\documentclass[pss,fleqn]{w-art}
\usepackage{times}
\usepackage{w-thm}
\usepackage{amssymb}

\usepackage[]{graphicx}
\chardef\bslash=`\\ 

\newcommand{\be}{\begin{equation}}
\newcommand{\ee}{\end{equation}}
\newcommand{\bea}{\begin{eqnarray}}
\newcommand{\eea}{\end{eqnarray}}
\newcommand{\br}{{\bf r}}

\newcommand{\bq}{{\bf q}}

\newcommand{\ve}{\varepsilon}
\newcommand{\e}{\epsilon}

\begin{document}
\DOIsuffix{theDOIsuffix}
\Volume{XX}
\Issue{X}
\Month{XX}
\Year{2007}
\pagespan{1}{}


\title[Magnetotransport in quantum Hall systems]{Magnetotransport of electrons
in quantum Hall systems}


\author[I.A. Dmitriev]{I.A. Dmitriev\inst{1}$^{,*}$}
\author[F. Evers]{F. Evers\inst{1}}
\author[I.V. Gornyi]{I.V. Gornyi\inst{1}$^{,*}$}
\author[A.D. Mirlin]{A.D. Mirlin\inst{1,2}$^{,\dagger}$}
\author[D.G. Polyakov]{D.G. Polyakov\inst{1}}
\author[P. W\"olfle]{P. W\"olfle\inst{2,1}}

\address[\inst{1}]{Institut
f\"ur Nanotechnologie, Forschungszentrum Karlsruhe, 76021 Karlsruhe,
Germany}
%
%
\address[\inst{2}]{Institut f\"ur Theorie der Kondensierten Materie,
Universit\"at Karlsruhe, 76128 Karlsruhe, Germany}
%

\thanks{$^*$ Also at A.F.~Ioffe Physico-Technical
Institute, 194021 St.~Petersburg, Russia\\
\phantom{aaaaa}\!$^\dagger$ Also at St.Petersburg Nuclear Physics Institute,
188300 St.Petersburg, Russia.}

\begin{abstract}
Recent theoretical results on magnetotransport of electrons in a 2D system in the range of
moderately strong transverse magnetic fields are reviewed. The
phenomena discussed include: quasiclassical memory effects in systems
with various types of disorder, transport in lateral superlattices,
interaction-induced quantum magnetoresistance, quantum
magnetooscillations in {\it dc} and {\it ac} transport, 
and oscillatory microwave photoconductivity.  
\end{abstract}
\maketitle                   




 \renewcommand{\leftmark}
 {I.A. Dmitriev et al.: Magnetotransport in quantum-Hall systems}


\section{Introduction}
\label{sec1}

Electronic transport in semiconductor nanostructures is one of 
the central issues of research in  modern condensed matter
physics, see, {\it e.g.,} \cite{beenakker91,ferry} for reviews.
In this article, we review recent results on transport of
two-dimensional electron gases (2DEG) in moderately strong transverse
magnetic fields $B$. Specifically, we concentrate on a range of $B$ which
are classically strong (i.e., $\omega_c\tau_{\rm tr}\gg 1$, where
$\omega_c$ is the cyclotron frequency and $\tau_{\rm tr}$ the transport
relaxation time), but where quantum localization effects (and,
correspondingly, quantum Hall physics) are not developed yet. 
There exists a broad class of phenomena that lead to strong
magnetoresistivity $\rho_{xx}(B)$ in this range of  fields, in view of
the developed cyclotron motion. These phenomena are discussed in the present
review.

\section{Quasiclassical memory effects in magnetoresistance}
\label{sec2}

The recent interest in {\it quasiclassical} transport properties of a 2DEG has
been largely motivated by the experimental and practical importance of
high-mobility heterostructures, in which charged impurities are separated from
the 2DEG by a wide spacer. The correlation radius of disorder produced by the
impurities is usually much larger than the Fermi wave length of electrons and
transport in the 2DEG retains signatures of the underlying quasiclassical
dynamics of the particles. On the theoretical side, much of the interest has
been inspired by a variety of anomalous transport phenomena which, while being
essentially classical, cannot be described by Boltzmann-Drude kinetic
theory. The quasiclassical ``non-Boltzmann" phenomena in disordered electron
systems are due to correlations of scattering acts at the points where
quasiclassical paths self-intersect, which gives rise to {\it memory effects},
neglected in the conventional Boltzmann equation. In particular, the
non-Markovian kinetics yields a strong magnetoresistance (MR) and anomalies in
the ac response. The strength of the quasiclassical anomalies depends on the
ratio $d/l$, where $d$ is the correlation radius of disorder, $l$ the mean
free path, and grows with increasing $d$ as a power of this parameter. Since
quantum corrections (weak localization, Altshuler-Aronov corrections, etc.)
are governed by a different small parameter $1/k_Fl\ll 1$, where $k_F$ is the
Fermi wave vector, it is the long-range correlations of disorder with $k_Fd\gg
1$ that reveal the quasiclassical anomalies. In this section, we focus on
the quasiclassical memory effects in magnetotransport.

\subsection{Magnetoresistance of a 2DEG subject to smooth disorder}
\label{s2.1}

We begin by considering the MR of a 2DEG in the presence of Gaussian disorder
which is smooth on the scale of $k_F^{-1}$. First of all, we recall that there
exists a finite MR \cite{khveshchenko96,mirlin98} even within the
collision-integral approximation, the source of which is the bending of
quasiclassical trajectories by magnetic field $B$ on the scale of the
correlation radius $d$. This leads to a small negative MR,
$\Delta\rho_{xx}/\rho_0\sim -(d/R_c)^2$, where $\rho_0$ is the Drude
resistivity and $R_c$ is the cyclotron radius. Remarkably, the non-Markovian
kinetics gives rise to a much stronger positive MR \cite{mirlin99}, which may
even be much larger than unity.

To systematically treat the quasiclassical memory effects, the starting point
is the disorder-averaged expression for the conductivity tensor $\sigma$ in
terms of the exact Liouville operator $L$:
\be
\sigma = e^2\nu v_F^2 \!\int \! \frac{d\phi}{2\pi} \left\langle\!
\left(\begin{array}{c} \cos{\phi}\\ 
                        \sin{\phi}\end{array}\right)
L^{-1} \!
\left(\begin{array}{c} \cos{\phi}\\ 
                       \sin{\phi}\end{array}\right)^T\right\rangle~.
\label{1}
\ee 
Here $\nu$ is the density of states, $v_F$ the Fermi velocity, $\phi$ the
velocity angle on the Fermi surface. The operator $L=L_0+\delta L$ is given by
the sum of the free part $L_0=-i\omega+v_{F}{\bf n}\nabla+
\omega_c\partial_\phi$, where $\omega_c$ is the cyclotron frequency, ${\bf
n}=(\cos\phi,\sin\phi)$, and the part induced by a random scalar potential
$V({\bf r})$,
\be
\delta L=\delta v({\bf r}){\bf n}\nabla+[\nabla\delta
v({\bf r})](\hat{\bf z}\times{\bf n})\partial_\phi~, 
\label{2}
\ee
where $\delta v({\bf r})=v({\bf r})-v_F$ denotes a fluctuation of the local
Fermi velocity $v({\bf r})=[v_F^2-2V({\bf r})/m)]^{1/2}$ and $\hat{\bf z}$ is a unit
vector in $z$ direction. Expanding Eq.~(\ref{1}) in $\delta L$, averaging over
the disorder and resumming the series, the diagonal resistivity
$\rho_{xx}=2(-i\omega+M_{xx})/e^2\nu v_F^2$ is represented in terms of the
disorder-induced self-energy (``memory function") $M_{xx}$. 

The Drude result $M_{xx}^{(0)}=\tau_{\text{tr}}^{-1}$, where $\tau_{\text{tr}}$ is the momentum
relaxation time, follows as the first term in a perturbative expansion of the
self-energy in the strength of disorder, $M_{xx}^{(0)}=-\langle\delta L \;
L_0^{-1}\; \delta L \rangle$. Substituting the propagator $L_{\rm D}^{-1}$
renormalized by impurity scattering for $L_0^{-1}$ in the latter expression
yields the main contribution to the MR associated with the quasiclassical
memory effects. When calculating $L_{\rm D}^{-1}$ in the case of long-range
disorder, the stochastic motion of particles can be approximated by a
Fokker-Planck equation corresponding to the diffusion in momentum space, so
that $L_{\rm D}$ is written as
\be
L_{\rm D}=-i\omega+v_F{\bf n}\nabla
+\omega_c\partial_\phi-\tau_{\text{tr}}^{-1}\partial^2_\phi~.
\label{3}
\ee
The leading correction to $M_{xx}$ due to the self-intersection of
quasiclassical paths then reads:
\be \Delta M_{xx}=( 4 \pi^3m^2v_F^2 )^{-1}\textstyle{\int}\!
d^2q\,d\phi\,\sin\phi\,
\sin(\phi-\phi_q)q^2W(q)\,g_{\rm D}(\omega,{\bf q},\phi)~, 
\label{4}
\ee
where $g_{\rm D}(\omega,{\bf q},\phi)$ is the Fourier-transformed real-space
solution of the equation $L_{\rm D}g_{\rm D}=\sin\phi\sin
(\phi-\phi_q)\delta(\bf r)$, $\phi_q$ is the angle of $\bf q$, and $W(q)$ is
the Fourier transform of the correlator $\left<V(0)V({\bf r})\right>$. For
the case of impurities separated from the 2DEG by a spacer of width $d$,
Eq.~(\ref{4}) gives \cite{mirlin99}
\be
\Delta\rho_{xx}/\rho_0=\Delta M_{xx}/M_{xx}^{(0)}=
2\pi^{-1}\zeta(3/2) \left(d/ l\right)^3 (\omega_c\tau_{\text{tr}})^{9/2}~.
\label{5}
\ee 

One sees that the MR due to the memory effects is much larger than that
related to the effect of magnetic field on the collision integral for
$\omega_c\tau_{\text{tr}}\gg (l/d)^{2/5}$. The MR (\ref{5}) becomes of order unity when
the mean-square shift of the guiding center of a cyclotron orbit after one revolution,
\be
\delta=2\pi^{1/2}v_F\tau_{\text{tr}}/(\omega_c\tau_{\text{tr}})^{3/2},
\label{5a}
\ee
becomes of order $d$, which happens at $\omega_c\tau_{\text{tr}}\sim (l/d)^{2/3}$. At higher fields,
the strong positive MR is followed by a sharp (exponential) falloff of
$\rho_{xx}$ with growing $B$ \cite{fogler97}:
\be
\ln (\rho_{xx}/\rho_0)\sim -(d/\delta)^{2/3}~,
\label{6}
\ee
which is due to the increasing adiabaticity of the electron dynamics and the
related quasiclassical localization. The self-intersection induced MR, given
by Eq.~(\ref{5}), may be considered as a precursor of the adiabatic
localization.

In the limit of weak inhomogeneities, the return-induced MR depends in an
essential way on the behavior of the disorder under time reversal
\cite{mirlin99}; in particular, it is strongly enhanced in a random magnetic
field (RMF). The case of a smoothly varying RMF is of particular interest in
view of the composite-fermion description of the transport properties of a
half-filled Landau level \cite{halperin93}. Also, a long-range RMF has been
realized in semiconductor heterostructures by attaching superconducting or
ferromagnetic overlayers or by ``prepatterning" the sample (randomly curving
the 2DEG layer). Following the same route as for the case of a random scalar
potential, the MR due to the quasiclassical memory effects is obtained as
\cite{mirlin99}
\be
\rho_{xx}/\rho_0=1/2+[1/4+(B/B_0)^2]^{1/2}~,
\label{7}
\ee     
where $B_0$ is the characteristic amplitude of fluctuations of the RMF. For
the composite-fermion model at half-filling, $(B/B_0)^2$ is represented as
$2(d/l)(\omega_c\tau_{\text{tr}})^2$. The adiabatic localization in the RMF begins at
$B\sim B_0(l/d)^{1/6}$ \cite{evers99}, so that there is a wide range of $B$ in
which the positive MR (\ref{7}) is strong.

\begin{figure}[htb]
\begin{minipage}[t]{.49\textwidth}
\includegraphics[width=0.8\textwidth]{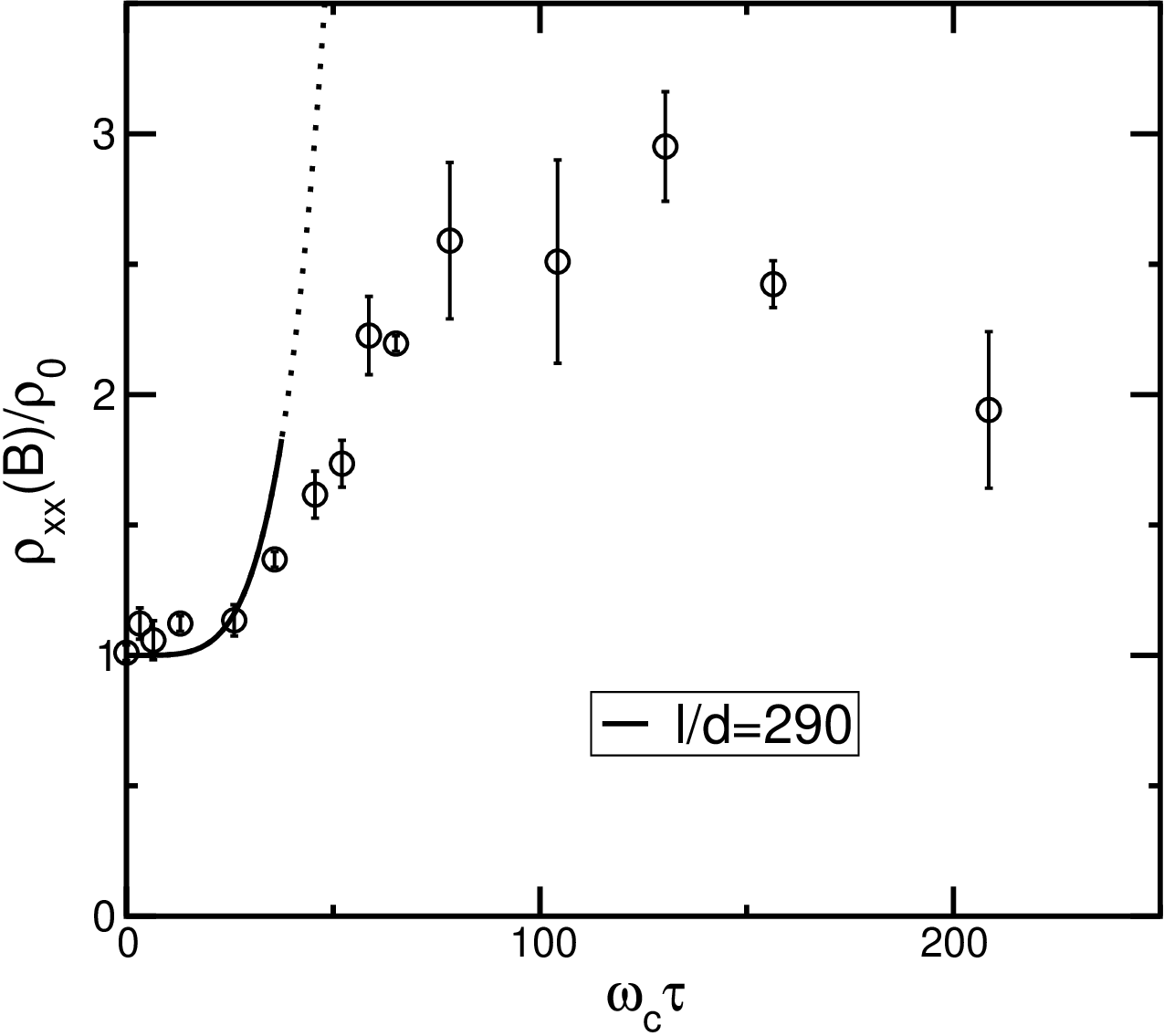}
\caption{
Magnetoresistivity in a random potential from numerical simulations in
comparison with Eq.~(\ref{5}) for $l/d=290$.}
\label{f1}
\end{minipage}
\hfill
\begin{minipage}[t]{.49\textwidth}
\includegraphics[width=0.8\textwidth]{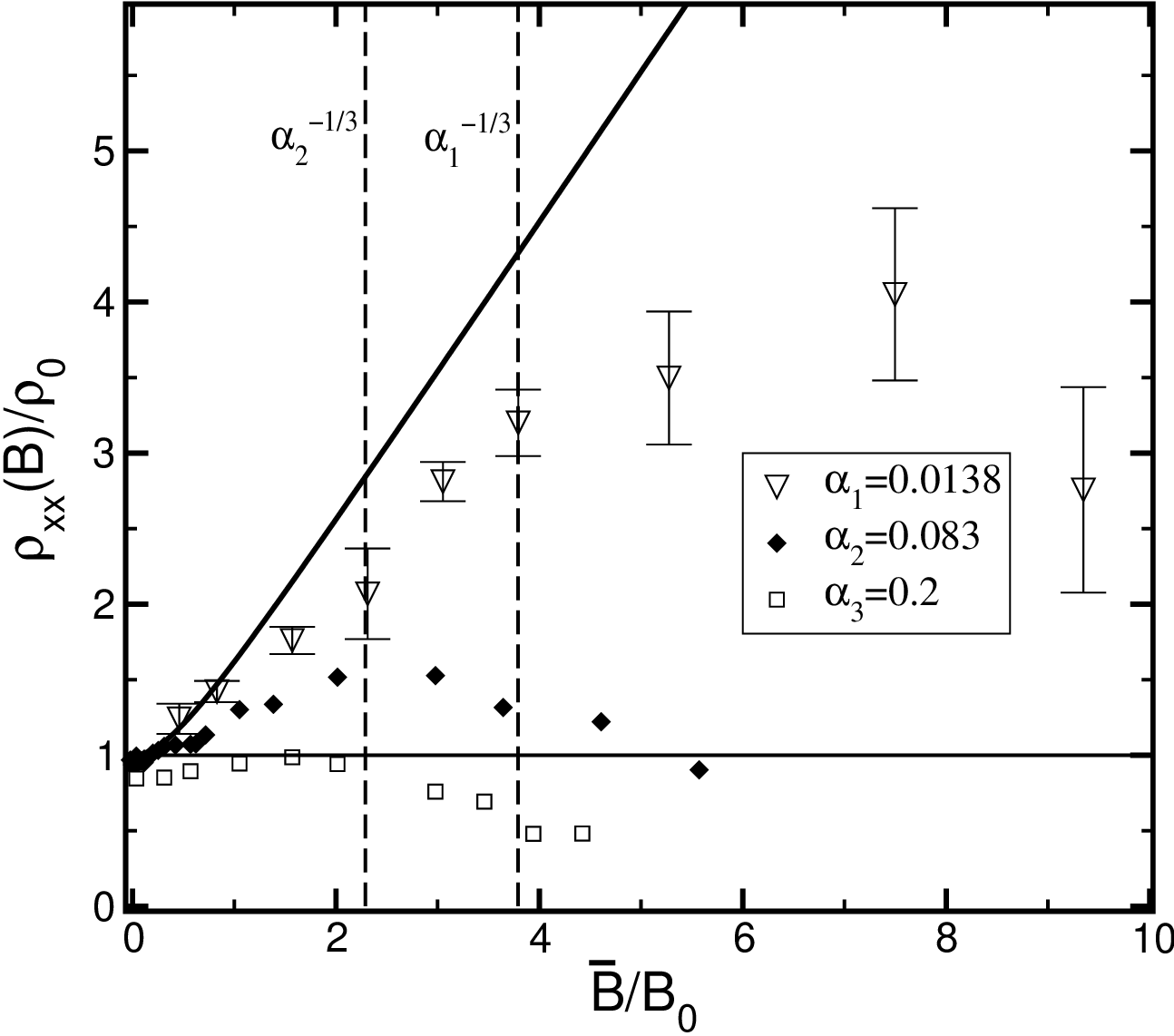} 
\caption{
Magnetoresistivity in a random magnetic field from numerical simulations
for three different strengths of the disorder $\alpha=(eB_0/mc)(d/v_F)$; the 
full line corresponds to Eq.~(\ref{7}).}
\label{f2}
\end{minipage}
\end{figure}

The numerically calculated MR \cite{mirlin99} for both types of disorder
shown in Figs.~\ref{f1} and \ref{f2} confirm the theoretically predicted
positive MR. Note that the MR in the RMF at
moderately small $d/l$ still exists, but becomes weak; this is the region of
$d/l$ relevant to the composite-fermion model. The numerical data for $d/l\sim
0.1-0.2$ agree well \cite{evers99} with the experimental results \cite{seesmet98}
for the MR around half-filling. Recent unpublished numerical simulations \cite{jullien}
also confirm the $\omega_c^{9/2}$ positive MR in a smooth potential; 
however, the numerical coefficient is found \cite{jullien} 
to be smaller by a factor $\sim 2$ as compared to Eq. ~(\ref{5}).

\subsection{Magnetoresistance of a 2DEG subject to two-component disorder}
\label{s2.2}

We now turn to a 2DEG moving in a non-Gaussian random potential represented by
rare strong short-range scatterers and subject additionally to a smooth random
potential discussed in Sec.~\ref{s2.1}. One of the most relevant experimental
realizations of the model of ``two-component disorder" is random antidot (AD)
arrays, where the potential barriers around the ADs can be modeled as hard
disks reflecting electrons specularly (for experimental work on the dc MR in
random AD arrays see, e.g.,
Refs.~\cite{gusev94,tsukagoshi95,luetjering96,nachtwei98,cina99,
yevtushenko00}). The model is also applicable to the description of the MR in
an unstructured ultra-high mobility 2DEG with a wide spacer, where large-angle
scattering on residual interface impurities and interface roughness becomes
important \cite{coleridge91,saku96,umansky97}, limiting the mobility with
further increasing width of the spacer. From the theoretical point of view,
the interplay of the two types of inhomogeneities is quite remarkable in that
it yields nontrivial physics which is absent in the limiting cases, when only
one type is present. In particular, although in the extreme of strong
$B\to\infty$ the resistivity $\rho_{xx}$ tends to zero in either of the
limiting cases, it diverges in the presence of both types of disorder
\cite{polyakov01}. Also, in the experimentally relevant situation when the
mean free path at zero $B$ is determined by scattering on ADs, the presence of
weak long-range disorder will nonetheless become of crucial importance with
increasing $B$ \cite{mirlin01,polyakov01}. The magnetotransport 
in the Lorentz-gas model describing an AD array without smooth disorder has 
been studied, in particular, in Refs.~\cite{entin,hauge,kuzmany,ddj,corridor}.
A strong negative MR followed by a metal-insulator phase transition 
in a strong magnetic field was found in \cite{entin,hauge,ddj}; the low-field anomalous MR due to the 
``corridor effect" was investigated in \cite{corridor}.
Smooth disorder, however, changes the MR qualitatively, as discussed below.

Generalizing the formalism described in Sec.~\ref{s2.1} to the case of
two-scale disorder, we represent the Liouville operator as $L=L_{\rm D}+\delta
L$, where $L_{\rm D}$ is given by Eq.~(\ref{3}) and includes interaction with
the smooth disorder ($\tau_{\text{sm}}$ is the corresponding transport scattering
time), whereas $\delta L=-\sum_i I_{{\bf R}_i}$ describes collisions with ADs
whose random positions are ${\bf R}_i$. The Fourier transform ${\tilde I}_q$
of the collision operator $I_{{\bf R}_i}$ yields the transport time
$\tau_S$ for the scattering by the AD array of density $n_S$ through ${\tilde
I}_0{\bf n}=-{\bf n}/n_S\tau_S$. We assume that
$\omega_c^{-1}\ll\tau_S\ll\tau_{\text{sm}}$, so that the total transport rate is determined by
ADs, $\tau_{\text{tr}}^{-1}=\tau_S^{-1}+\tau_{\text{sm}}^{-1}\simeq \tau_S^{-1}$. 
The leading contribution to $\Delta M_{xx}$
reads
\be
\Delta M_{xx}=-n_S\textstyle{\int} (d\phi/\pi)\cos\phi\,I_{\bf R}DI_{\bf R}\,\cos\phi~,
\label{8}
\ee 
where the propagator $D$ includes the first-order self-energy: $D=(L_{\rm
D}-n_S{\tilde I}_0)^{-1}$. As compared to the Lorentz model \cite{ernst71} in
which only hard-disk scatterers are present, new physics emerges in the limit
$\delta\gg a$, where $a$ is the radius of the ADs and $\delta$ is the shift of
the cyclotron orbit after one revolution due to scattering on smooth disorder
(cf.\ Sec.~{\ref{s2.1}). In particular, Eq.~(\ref{8}) yields \cite{mirlin01}
\be
\Delta\rho_{xx}/\rho_0=-(\omega_c/\omega_0)^2~,\quad \omega_0
=(2\pi n_S)^{1/2}v_F(3\tau_S/2\tau_{\text{sm}})^{1/4}~,
\label{9}
\ee
for $\omega_c\ll\omega_0$. The mechanism of the negative MR (\ref{9}) can be
understood as follows. If one associates with a particle trajectory a strip of
width $2a$, the ratio $(\omega_c/\omega_0)^2$ gives the fraction of the area
``explored" twice, which implies an effective reduction of the exploration
rate and thus a longer time between collisions with {\it different} ADs. The
negative MR (\ref{9}) should be contrasted with the positive MR (\ref{5}) for
one-scale smooth disorder, where the passages through the same area lead to an
enhanced scattering rate.

For $\omega_c\gg\omega_0$, the renormalized scattering time $\tau'_S\gg\tau_S$
should be found self-consistently from the condition $n_S\xi R_c\sim 1$, where
$\xi\sim \delta(v_F\tau'_S/R_c)^{1/2}$ is a characteristic end-to-end size of
the diffusive guiding-center trajectory in time $\tau'_S$, which gives the 
$B^{-4}$ falloff \cite{mirlin01} with increasing $B$:
\be
\rho_{xx}/\rho_0\sim\tau_S/\tau'_S\sim (\tau_S/\tau_{\text{sm}})(n_SR_c^2)^2~.
\label{10}
\ee 
Equation (\ref{10}) is valid as long as $\tau'_S\ll\tau_{\text{sm}}$, which is rewritten
as $n_SR_c^2\gg 1$. In the opposite limit (but still for $\delta\ll d$), the
scattering on ADs stops playing any role and $\rho_{xx}$ has a plateau with
\be
\rho_{xx}/\rho_0=\tau_S/\tau_{\text{sm}}~.
\label{11}
\ee

Let us now briefly outline what happens at larger $B$, namely for $\delta/d\ll
1$ \cite{polyakov01}. In dilute AD arrays for intermediate $B$, the
exponential slowdown of the electron dynamics induced by adiabatic
localization transforms into $\rho_{xx}/\rho_0\sim
n_SR_cd\ln(1/n_SR_cd)\propto B^{-1}\ln B$ due to rare collisions with ADs
which mix otherwise closed drift trajectories in phase space. Most
interesting, however, is the behavior of $\rho_{xx}$ in the limit of large
$B$, where $\rho_{xx}$ starts to grow as a power law with increasing $B$. This
behavior can be most clearly seen in a ``hydrodynamic model" of the chaotic AD
array ($n_S\to\infty$, $\tau_S={\rm const}$), where the problem can be mapped
onto that of advection-diffusion transport \cite{isichenko92}. In the limit
$B\to\infty$ the hydrodynamic model predicts $\rho_{xx}/\rho_0\sim
(\tau_S^2v_Fd/\tau_{\text{sm}} R_c^2)^{5/13} \propto B^{10/13}$. The physics of the
positive MR is a percolation of drifting cyclotron orbits limited by
scattering on ADs. The growth is checked on the side of large $B$ by quantum
effects (Shubnikov-de Haas oscillations). The different types of the MR in the
two-component model \cite{polyakov01} 
are illustrated in Fig.~\ref{sketchy} for various
concentrations of short-range scatterers (ADs): (i)
the MR is positive owing to the ``diffusion-controlled
percolation"; (ii) due to the adiabatic localization, the
concentration of conducting electrons decreases as $B^{-1}\ln B$
before the percolation becomes effective, which yields a negative
MR $\rho_{xx}(B)\propto B^{-1}\ln B$ for intermediate
$B$; (iii) an exponentially sharp falloff of $\rho_{xx}(B)$ at $B\sim
B_{ad}$ (shown as a vertical jump) separates the diffusive and drift
regimes; (iv) because of the memory effects, the collision time for
scattering by ADs is increased as compared to the Drude value
already in the diffusive regime ($B\ll B_{ad}$), which leads to the
negative MR $\rho_{xx}(B)\propto B^{-4}$ for small $B$;
(v) for intermediate $B$, the scattering on ADs stops playing any
role and $\rho_{xx}(B)$ is saturated at a value determined by the
long-range disorder only, whereas at larger fields the
diffusion-controlled percolation gives rise to a positive
MR. 

\begin{figure}[htb]
\begin{minipage}[t]{.48\textwidth}
\includegraphics[width=0.8\textwidth]{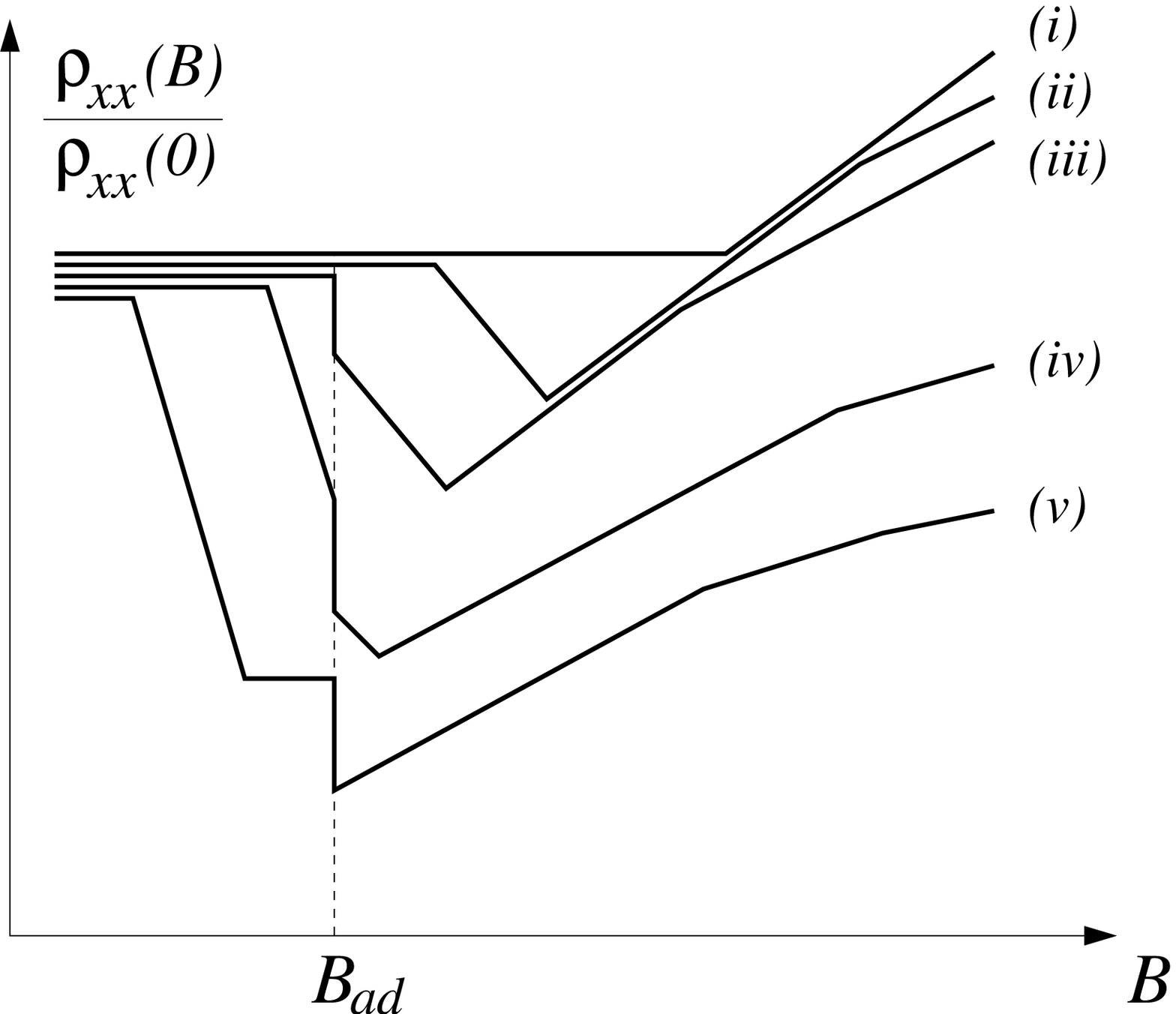} 
\caption{
Schematic behavior of the magnetoresistivity $\rho_{xx}(B)$
on a log-log scale in the two-component disorder model 
for different values of the concentration of
antidots $n$: $n^{(i)}>n^{(ii)}>\ldots >n^{(v)}$, keeping all other
parameters ($\tau_S,\tau_{\rm sm},d$) fixed. Only one characteristic field 
$B_{ad}$
is shown, at which the crossover between diffusive dynamics and
adiabatic drift in the long-range potential takes place.}
\label{sketchy}
\end{minipage}
\hfill
\begin{minipage}[t]{.48\textwidth}
\includegraphics[width=0.8\textwidth]{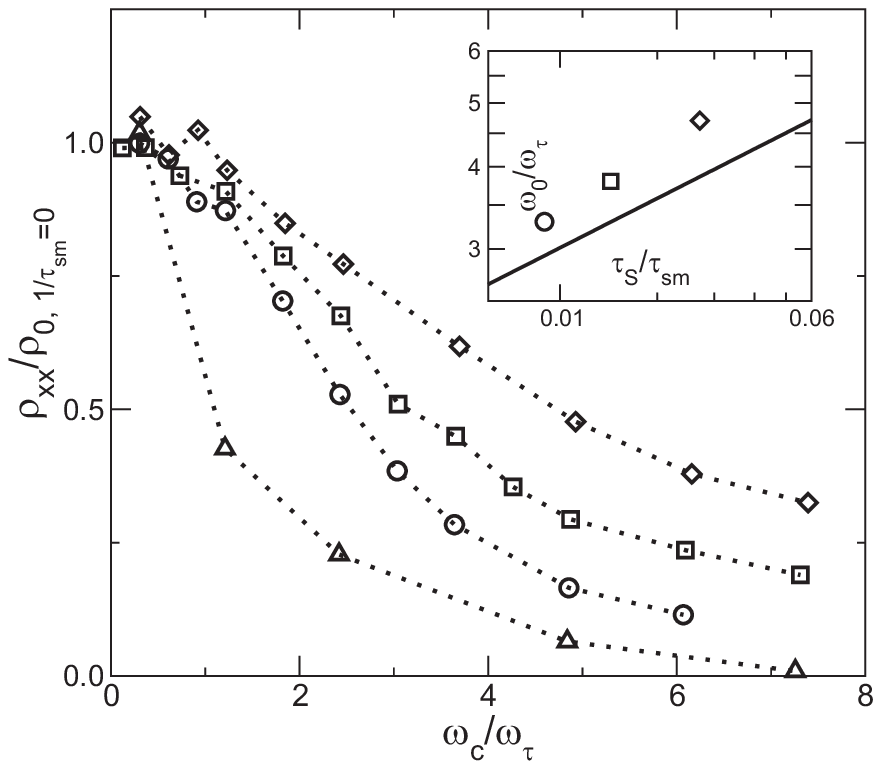}
\caption{ 
Magnetoresistivity in a two-component-disorder model 
at fixed $\tau_S$ and different $\tau_{\text{sm}}$ from
numerical simulations; $\tau_{\text{sm}}/\tau_S=\infty$ (Lorentz gas, 
$\triangle$), 111
($\bigcirc$), 70 ($\square$), 
37 ($\Diamond$). 
Inset: $\omega_0$ determined from the fit to the quadratic behavior given by
Eq.~(\ref{9}); the full line corresponds to the analytical result for
$\omega_0$ from Eq.~(\ref{9}).
}
\label{f3}
\end{minipage}
\end{figure}

The MR obtained numerically \cite{mirlin01} for the model of two-scale
disorder in the regime of Eqs.~(\ref{9})-(\ref{11}), Fig.~\ref{f3}, shows good
agreement with the above analytical results. As far as the experimental data
are concerned, for typical parameters \cite{luetjering96} of AD arrays,
$n_S=(0.6\,\mu {\rm m})^{-2}$, $v_F\tau_S=1.3\,\mu {\rm m}$, $v_F\tau_{\text{sm}}=16\,\mu
{\rm m}$, and the electron density $5\times 10^{11}\,{\rm cm}^{-2}$, the field
$B_0$ corresponding to the frequency $\omega_0$ given by Eq.~(\ref{9}) is
$\simeq 0.3\,T$, in agreement with the experimental findings
\cite{luetjering96}. A similar negative MR was reported in
Refs.~\cite{gusev94,tsukagoshi95,nachtwei98}. For ultra-high mobility samples
[electron density $2\times 10^{11}\,{\rm cm}^{-2}$, $v_F\tau_S\simeq
80\,\mu{\rm m}$, $\tau_{\text{sm}}/\tau_S\sim 10$, $a\sim 10\,{\rm nm}$, $n_S\sim
(2\,\mu{\rm m})^{-2}$ \cite{saku96}], one gets $B_0\sim 60\,{\rm mT}$. A
strong negative MR has indeed been observed \cite{umansky97,smetpriv} in the
very-high-mobility heterostructures, in qualitative agreement with the above
theory.

\subsection{Quasiclassical memory effects in ac magnetotransport}
\label{s2.3}

In addition to the strong MR, the non-Markovian quasiclassical kinetics in the
presence of long-range disorder gives rise to an anomalous ac response.  In
particular, the return-induced correction to the ac conductivity ${\rm
Re}\,\sigma(\omega)$ exhibits a kink \cite{wilke00} $\propto |\omega|$ at
$\omega\to 0$. The quasiclassical zero-frequency anomaly is not sensitive to
inelastic scattering (in contrast to the weak-localization quantum correction)
but manifests itself only \cite{evers01} in the presence of an external
metallic gate that screens the long-range Coulomb interaction. As outlined
below, in a strong $B$, the diagonal ac conductivity ${\rm
Re}\,\sigma_{xx}(\omega)$ shows pronounced resonant features
\cite{polyakov02,dmitriev04} on top of the cyclotron resonance (CR), which are
induced by the memory effects.

In the Lorentz model (hard disks with no smooth disorder), which is intended
to mimic a random AD array in heterostructures in the limit of a large spacer,
the shape of the CR is very different from the Lorentzian and not
characterized by the Drude scattering rate \cite{kuzmany,polyakov02}. 
Altogether, the behavior of ${\rm
Re}\,\sigma_{xx}(\omega)$ associated with quasiclassical cyclotron orbits
skipping around the ADs turns out to be remarkably rich \cite{polyakov02}. The
skipping-orbit contribution is broadened on a scale of $\omega_c$ and vanishes
at $\omega_c$ in a nonanalytical way as $|\omega-\omega_c|$. Apart from these
two features, ${\rm Re}\,\sigma_{xx}(\omega)$ for moderately strong $B$ with
$R_c\gg a$ oscillates with a period $\omega_c$ up to $\omega=\omega_cR_c/a$
and shows a series of square-root spikes for larger $\omega$. The modulation
yields exact zeros of the ac response at the harmonics of the CR.

Adding a smooth random potential, present in typical heterostructures, changes
the above picture in an essential way. Following the formalism of
Sec.~\ref{s2.2}, the classical return-induced contribution to the real part of
the oscillatory ac conductivity $\sigma^{(c)}_\omega$ is represented for
$\tau_S\ll\tau_{\text{sm}}$ and $\delta\gg d$ as \cite{dmitriev04}
\be 
\sigma^{(c)}_\omega /\sigma^{\rm D}_\omega=-{\rm
Re}\,P_\omega/n_S\tau_S~,\qquad P_\omega=\textstyle{\sum}_{n=1}^\infty \textstyle{\int} \! dt \,
e^{-i\omega t-t/\tau_S}\,p_n[\,0,v_F(t-2\pi n/\omega_c)\,]~,
\label{12} 
\ee
where $\sigma^{\rm D}_\omega$ is the real part of the dynamic Drude
conductivity, Eq.~(\ref{drude}).
In Eq.~(\ref{12}), the function $p_n(x_\perp,x_\parallel)=(3^{1/2}\pi
n^2\delta^2)^{-1}\exp\left[-(3x_\perp^2+x_\parallel^2)/ 3n\delta^2\right]$
describes the electron distribution in time $t=2\pi n/\omega_c$ after $n$
cyclotron revolutions along $(x_\parallel)$ and across $(x_\perp)$ the
cyclotron orbit. In the case of a weak damping of the quasiclassical
magnetooscillations, $\sigma^{(c)}_\omega$ is represented as a series of sharp
dips at $\omega=N\omega_c$, whose amplitude and width are of order
$\sigma^{\rm D}(a/N\delta)(\omega_c\tau)^{1/2}$ and
$\tau_S^{-1}+N^2\tau_{\text{sm}}^{-1}$, respectively. In the regime of harmonic
oscillations (exponential damping), $\sigma_\omega^{(c)}$ is written as 
\cite{dmitriev04}
\be
\sigma_\omega^{(c)}/\sigma_\omega^{\rm D}
=1-(a/\pi^{1/2}\delta)
\,\cos(2\pi\omega/\omega_c)
\,\exp\left[-(\omega/\omega_c)^2(3\pi/\omega_c\tau_{\text{sm}})\right]~.
\label{14}
\ee
It is worth stressing that these oscillations are of essentially classical
origin and have nothing to do with the Landau quantization. The behavior of
$\sigma_\omega^{(c)}$ is illustrated in Fig.~\ref{f4}, where also the quantum
oscillations $\propto \exp(-2\pi/\omega_c\tau_q)$, with $\tau_q$ being the
single-particle (quantum) relaxation time, are shown. One sees that the
classical oscillations may be stronger than the quantum ones since in
high-mobility structures $\tau_q\ll\tau_{\rm sm}$ and the quantum oscillations are
damped much more strongly.

\begin{SCfigure}[8][htb]
\includegraphics[width=.4\textwidth]{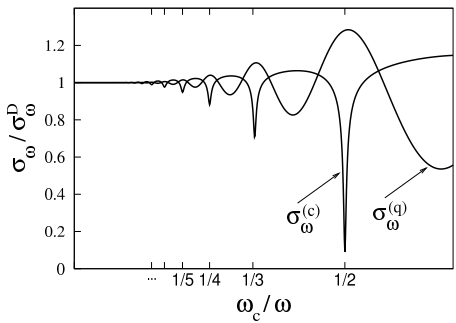}
\caption{
Quasiclassical [$\sigma_\omega^{(c)}$, Eq.~(\ref{14})] and quantum
[$\sigma_\omega^{(q)}$] oscillatory ac conductivity (normalized to the Drude
conductivity $\sigma_\omega^{\rm D}$) vs $\omega_c/\omega$ for
$\omega/2\pi=100$~GHz, $\tau_{\rm sm}=0.6$~ns, $\tau_{\rm sm}/\tau_{\rm q}=50$,
$\tau_S/\tau_{\rm sm}=0.1$, $a/\delta=0.25$ at $\omega_c/\omega=1/2$.
}
\label{f4}
\end{SCfigure}

\section{Magnetotransport in modulated systems (lateral superlattices)}
\label{s8}

\subsection{Weiss oscillations in one-dimensional superlattices}
\label{s8.1}

 Transport properties of a two-dimensional electron gas
(2DEG) subject to a periodic potential (lateral superlattice) with a
period much shorter than the electron transport mean free path (but
much larger than the Fermi wave length) have
been intensively studied during the last decade. In a pioneering
experiment \cite{weiss} Weiss
{\it et al.} discovered that a weak one-dimensional (1D) modulation 
with wave vector ${\bf q}\parallel {\bf e}_x$ induces strong
commensurability oscillations of the magnetoresistivity
$\rho_{xx}(B)$ (while showing almost no effect on $\rho_{yy}(B)$ and
$\rho_{xy}(B)$), with the minima satisfying the
condition $2R_c/a=n-1/4$, $n=1,2,\ldots$, where $R_c$ is the
cyclotron radius and $a=2\pi/q$ the modulation wave
length. The quasiclassical nature of these commensurability oscillations
was demonstrated by Beenakker \cite{beenakker}, who showed that the
interplay of the cyclotron motion and the superlattice potential
induces a drift of the guiding center along $y$ axis, with an
amplitude squared oscillating as
$\cos^2(qR_c-\pi/4)$ (this is also reproduced by a quantum-mechanical
calculation, see \cite{theory-qm}). While describing nicely the period
and the phase of the experimentally observed oscillations, the result
of \cite{beenakker}, however, failed to explain the observed rapid
decay of the oscillation 
amplitude with decreasing magnetic field. The cause for this
discrepancy was in the treatment of disorder: while 
Ref.~\cite{beenakker} assumed isotropic impurity scattering, in
experimentally relevant high-mobility semiconductor heterostructures
the random potential is 
very smooth and induces predominantly small-angle scattering, with
the total relaxation rate $\tau_{\rm q}^{-1}$ much exceeding the momentum
relaxation rate $\tau_{\rm tr}^{-1}$.
The theory of commensurability oscillations in one-dimensional
modulation, $V(x)=\eta E_F\cos qx$ with $\eta\ll 1$,
in the situation of smooth disorder was worked out in \cite{mw98}. 

The starting point is the Boltzmann equation
for the distribution function $F(x,{\bf n})$ of electrons,
\begin{equation}
{\cal L} F(x,{\bf n}) = -e v(x) {\bf En}\ ; \qquad
{\cal L}=v(x){\bf n}\partial_{\bf r} +
\omega_c \partial_\phi-\sin\phi
v'(x) \partial_\phi - C\ , \label{e2}
\end{equation}
where ${\bf n}=(\cos\phi,\sin\phi)$ is the direction and
 $v(x)=[2m(E_F+eU(x))]^{1/2}$ the magnitude of the Fermi
velocity, and $C$ is the collision
integral. The resulting modulation-induced contribution to resistivity
reads \cite{mw98}
\begin{equation}
\frac{\Delta\rho_{xx}}{\rho_0}  =  \frac{\eta^2 ql}{4} Q
\frac{\pi}{\sinh\pi\mu} J_{i\mu}(Q)J_{-i\mu}(Q)\ ; \qquad
\mu=\frac{Q}{qv_F\tau_{\rm q}}
\left[1-\left(1+\frac{\tau_{\rm q}}{\tau_{\rm tr}}Q^2\right)^{-1/2}\right]\ ,
\label{e11}
\end{equation}
where we introduced the dimensionless parameter $Q=qR_c$ convenient to
characterize the strength of the magnetic field. At low 
magnetic fields, $Q\gg Q_{\rm dis}$, with $Q_{\rm dis}=(2ql/\pi)^{1/3}$, the
oscillations are exponentially damped by disorder, and the
magnetoresistivity saturates at the value
$\Delta\rho_{xx}/\rho_0  =  \eta^2 ql / 4$.
(At still lower
magnetic fields, $Q>Q_{\rm ch}$, with $Q_{\rm ch}=2/\eta$, and for a
sufficiently 
strong modulation, $\eta^{3/2} ql\gg 1$, an additional strong
magnetoresistivity occurs, dominated by the channeled orbits, see
Sec.~\ref{s8.3}.)
 In strong fields, $Q\ll Q_{\rm dis}$, the amplitude of oscillations
increases as $B^3$,
\begin{equation}
\Delta\rho_{xx} / \rho_0 =[(\eta ql)^2 /  \pi Q^3] \cos^2(Q-\pi/4)\ .
\label{e14}
\end{equation}

In Fig.~\ref{f8.1} the theoretical results are compared  
with experimental data of
Weiss {\it et al} \cite{weiss}. The sample parameters are \cite{weiss} 
$q=2\pi/382\mbox{nm}$, $n_e=3.16\times
10^{11}\mbox{cm}^{-2}$, $\tau_{\rm tr} = 52\mbox{ps}$, 
the total relaxation rate
is taken to be $\tau_{\rm q}^{-1}=(3\mbox{ps})^{-1}$.
As is seen from the figure, 
with the modulation strength  $\eta=0.065$ a very good description
of the experimentally observed magnetoresistivity is obtained. (At
lowest $B$, the experimental data show positive magnetoresistance
discussed in Sec.~\ref{s8.3}.)
The difference between  the long-range potential scattering
and the isotropic scattering is illustrated in the right panel,  
where the results for the modulation-induced $\Delta\rho_{xx}$
are plotted for both models of disorder 
at the same value of $\tau_{\rm tr}$.
As was shown in \cite{lewm98}, modulation-induced 
commensurability oscillations can
be also observed in attenuation and velocity change of a surface
acoustic wave propagating near a 2DEG.

\begin{figure}[htb]
\includegraphics[width=0.45\textwidth]{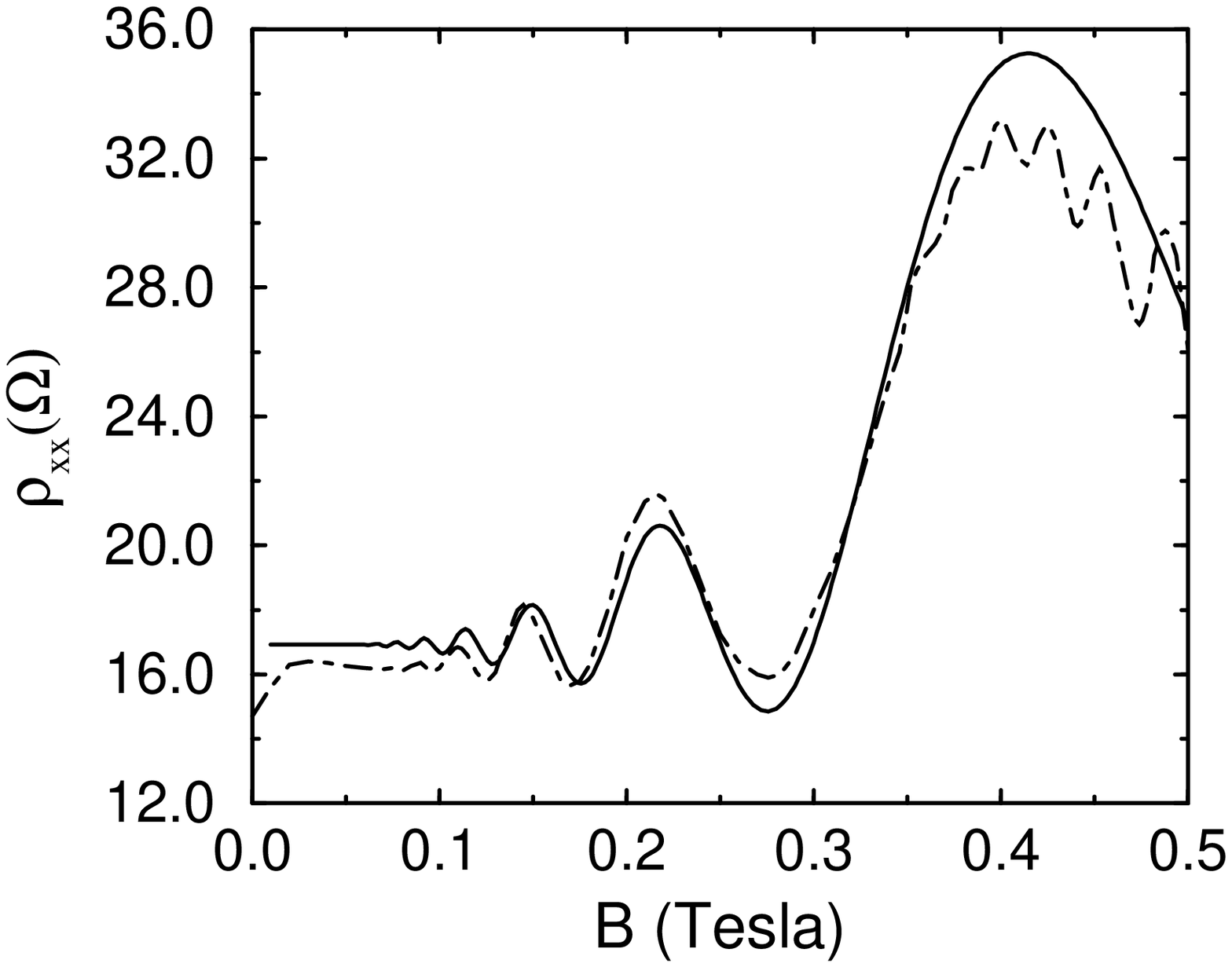} \hskip1cm
\includegraphics[width=0.45\textwidth]{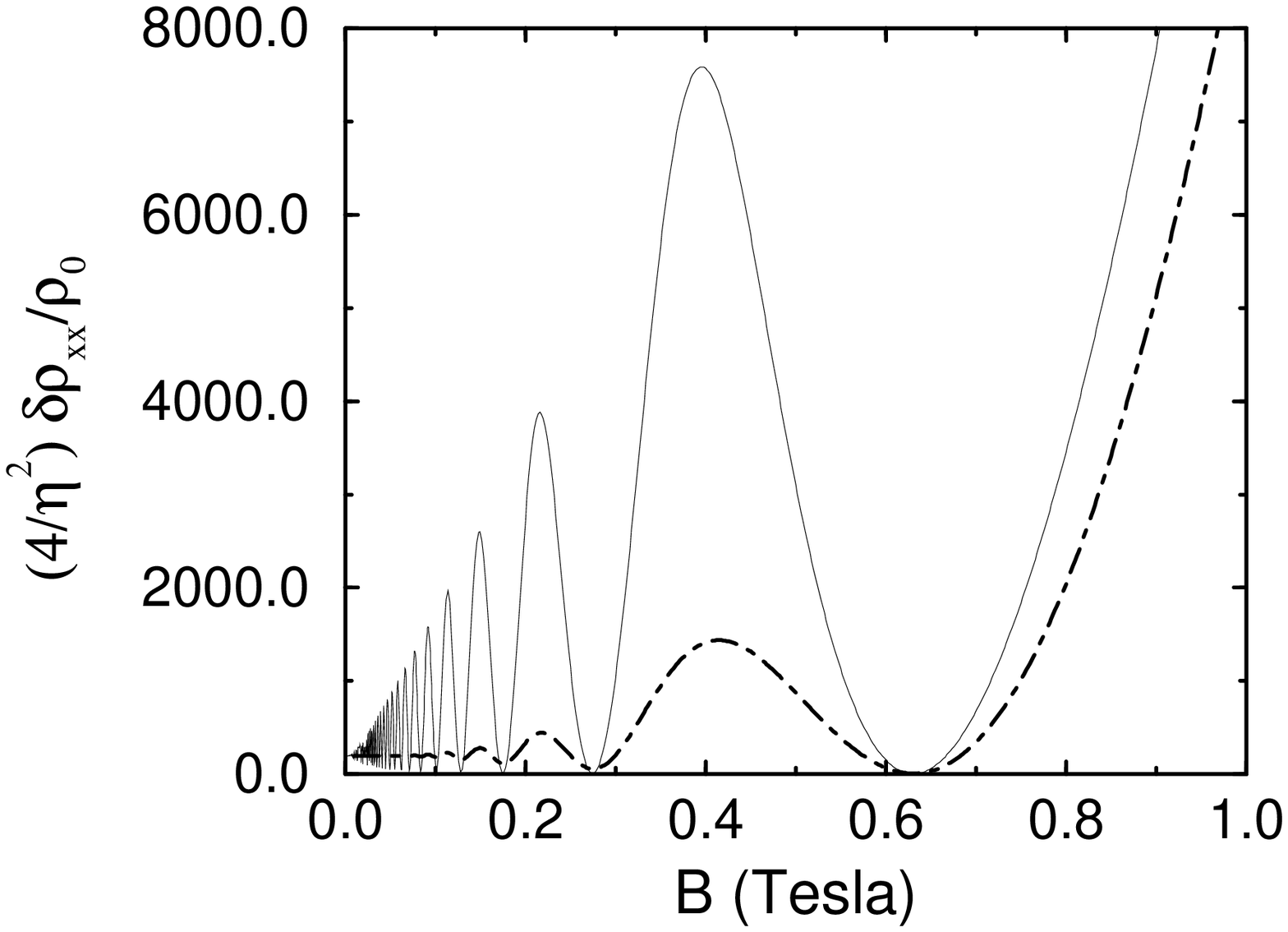}
\caption{Weiss oscillations in a 1D superlattice. 
{\it Left panel:}
Experimental data of Ref.\protect\cite{weiss} (dash-dotted line)
compared with the theoretical results for the long-range potential
scattering (full line). Parameters are:
$q=2\pi/382\mbox{nm}$, $n_e=3.16\times
10^{11}\mbox{cm}^{-2}$, $\tau=52\mbox{ps}$,
$\tau_s=3\mbox{ps}$, $\eta=0.065$. {\it Right panel:}
Grating-induced correction to the 2DEG reistivity,
$\Delta\rho_{xx}/\rho_0$, in units of $\eta^2/4$ for the isotropic
potential scattering ($\tau_{\rm tr}=52\mbox{ps}$, full line) and the
long-range random potential scattering ($\tau_{\rm tr}=52\mbox{ps}$,
$\tau_{\rm q}=3\mbox{ps}$, dash-dotted line). The sample parameters
($q=2\pi/382\mbox{nm}$, $n_e=3.16\times
10^{11}\mbox{cm}^{-2}$)   are the
same as in Ref.\protect\cite{weiss}.}
\label{f8.1}
\end{figure}

In \cite{mwle98} Weiss oscillations were studied for the vicintiy of
the $\nu=1/2$ filling of the lowest Landau level; see also related
works \cite{vonoppen98,zwerschke99}. Within the composite
fermion theory, the problem is described in terms of fermions subject
to a spatially modulated magnetic field and scattered by a random
magnetic field.  The magnetic character of modulation shifts the phase
of Weiss oscillations, while the random magnetic fields considerably
enhances their damping. The obtained results are in agreement with
experimental studies \cite{smet98,smet99}, which confirms the validity
of the composite-fermion description of the $\nu=1/2$ state.

\subsection{Two-dimensional superlattices}
\label{s8.2}

Magnetotransport in 2D superlattices with small-angle
impurity scattering was studied in \cite{mtw1}. It was shown that 
the shape of the magnetoresistivity depends crucially on the
parameter $\gamma = \eta^2ql/4$.  

For small $\gamma$ (corresponding typically to a
modulation strength not exceeding a few percent) the
magnetoresistivity is given by the perturbative formulas (\ref{e11})
-- (\ref{e14}) (the same as in 1D superlattices) 
up to the point $Q\sim Q_{\rm P} \equiv [0.13(\eta ql)^2]^{1/3}$, 
where the correction
$\Delta\rho_{xx}$ becomes of the order of the Drude resistivity
$\rho_0$. For higher magnetic fields the P\'eclet number $P\sim \eta
ql/(qR_c)^{3/2}$ 
characterizing the advection-diffusion problem becomes large and the
transport is determined by a narrow boundary layer around a square
network of separatrices. As a result, the $B^3$-dependence of the
oscillation amplitude characteristic for the perturbative 
($Q>Q_{\rm P}$) regime crosses over to a much slower
$B^{3/4}$-increase at $Q<Q_{\rm P}$, 
\be
\label{e17}
\Delta\rho_{xx} / \rho_0 =(8\pi)^{1/4}C(\eta ql)^{1/2}Q^{-3/4}
|\cos(Q-\pi/4)|^{1/2}\ .
\ee

For $\gamma\gg 1$ (which is typically valid for the modulation
strength $\eta$ larger than $10 \div 15 \%$) the oscillations are
damped at low magnetic fields not by disorder (as in the perturbative
regime) but by the modulation-induced chaotic diffusion. The
oscillations become observable at $Q\sim  Q_{\rm ad} \equiv
(4\pi/\eta^2)^{1/3}$  where the
motion of electrons in the superlattice potential acquires the form of
adiabatic drift. Since the violation of adiabaticity is exponentially
small, the magnetoresistivity drops exponentially in a logarithmically
narrow interval of magnetic fields, 
$Q'_{\rm ad} \equiv Q_{\rm ad}(\ln\gamma)^{-2/3}< Q<Q_{\rm ad}$, 
\be
\label{e23}
\Delta\rho_{xx} / \rho_0\propto 
\exp [-(\pi / 2\sqrt{2}|\cos(Q-\pi/4)|)
(Q_{\rm ad} /  Q)^{3/2}]\ .
\ee
At higher
magnetic fields the impurity scattering starts to dominate over the
non-adiabatic processes and thus to determine the diffusion constant of
the advection-diffusion problem, so that the commensurability
oscillations take the same form (\ref{e17}) as in the
large-$P$ limit of the $\gamma\ll 1$ regime.

\begin{figure}[htb]
\includegraphics[width=0.4\textwidth]{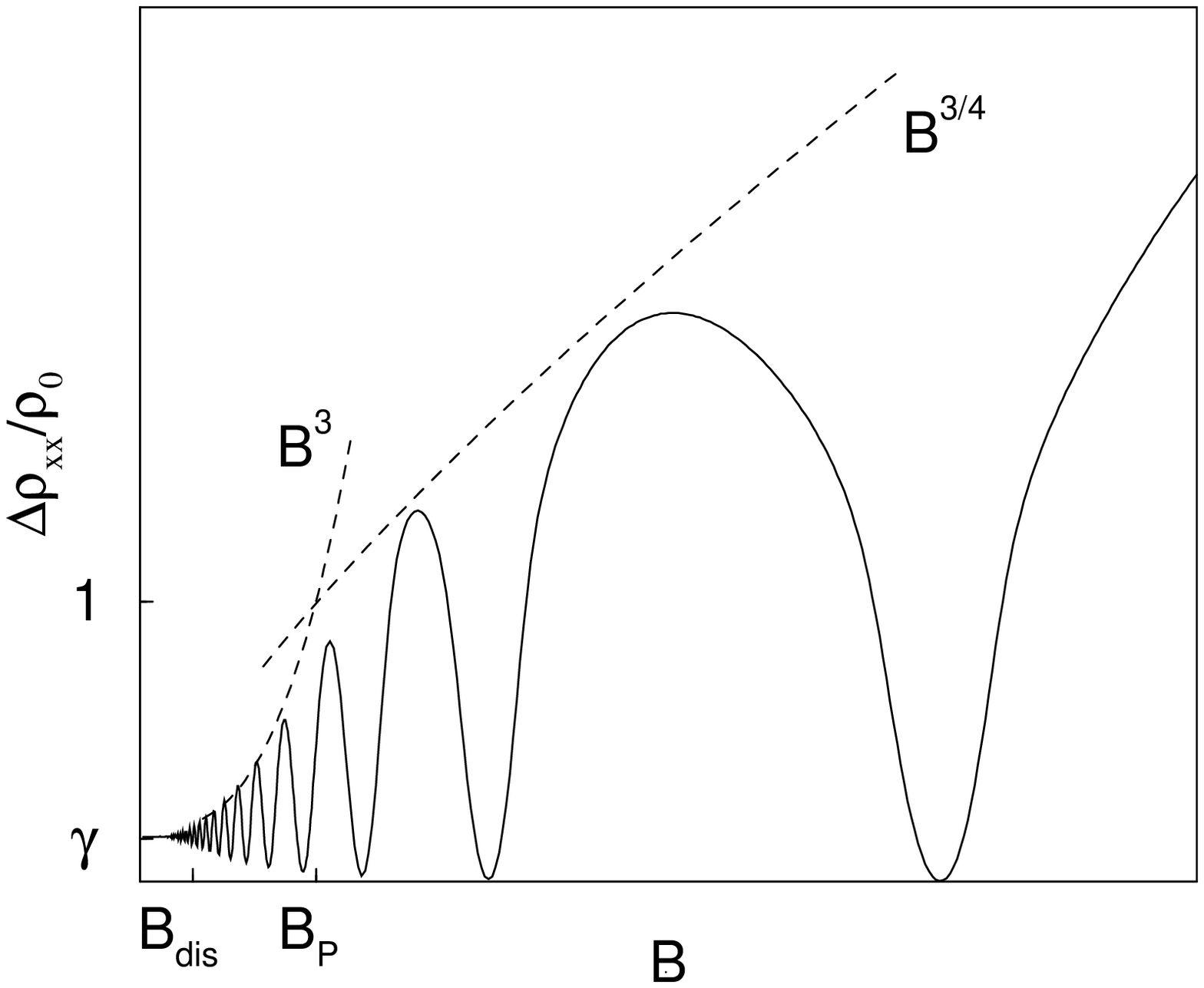} \hskip1cm
\includegraphics[width=0.42\textwidth]{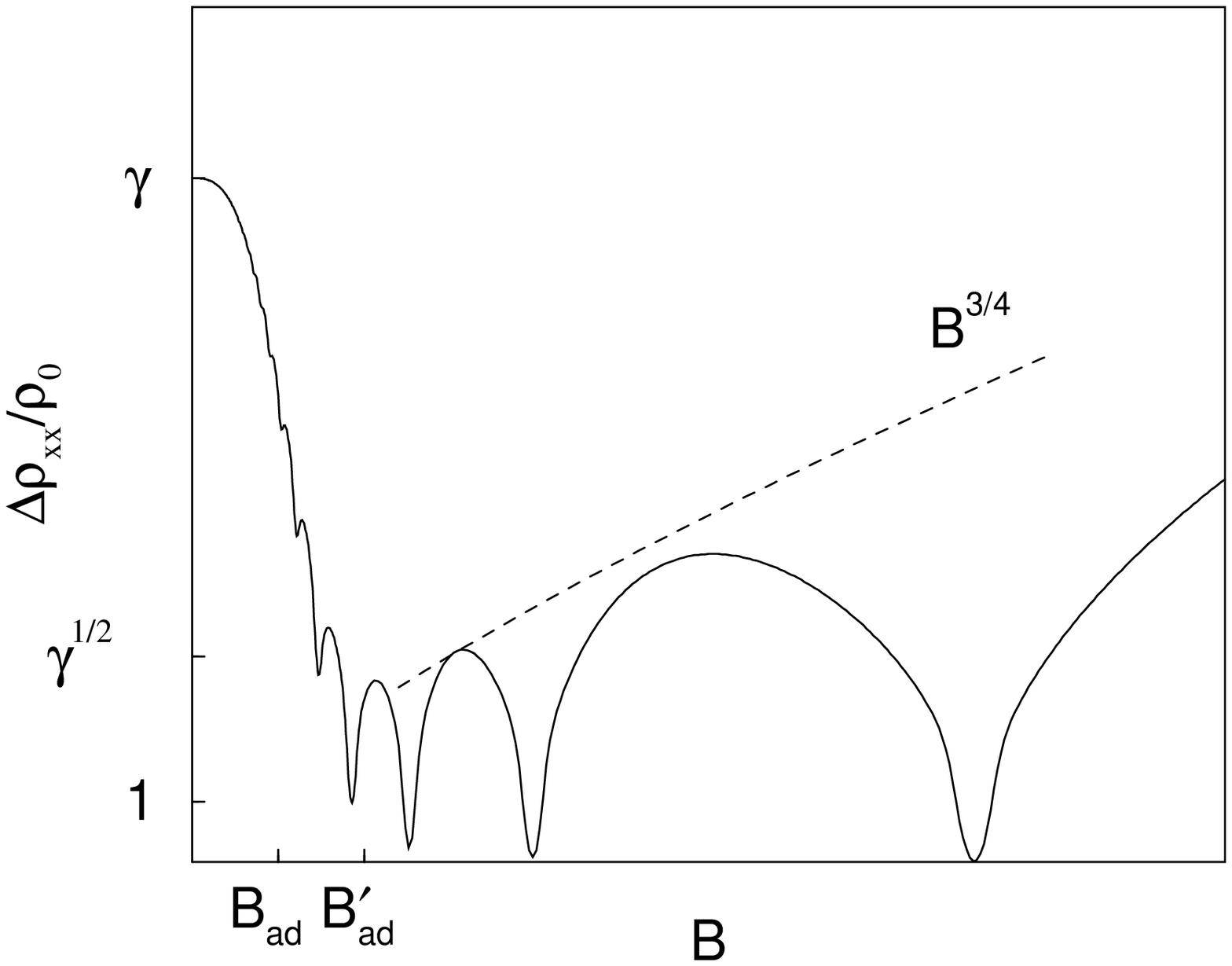}
\caption{ {\it Left panel:} Schematic representation of the magnetoresistivity
$\Delta\rho_{xx}(B)$ induced by a 2D modulation in the case $\gamma\ll
1$. Characteristic points $B_{\rm dis}$ and $B_P$ on the magnetic
field axis (corresponding to $Q=Q_{\rm dis}$ and $Q=Q_P$, see the text)
are shown. Below $B_{\rm dis}$ the oscillations are exponentially
damped and the magnetoresistivity saturates at
$\Delta\rho_{xx}=\gamma$, while at $B=B_{\rm P}$ the $B^3$-behavior of
$\Delta\rho_{xx}$ changes to a much slower, $B^{3/4}$-increase.
{\it Right panel:} Schematic representation of 
$\Delta\rho_{xx}(B)$ in the case $\gamma\gg 1$. 
The magnetoresistivity starts to drop exponentially and the
commensurability oscillation appear at the value $B_{\rm ad}$ of the
magnetic field where the motion in the periodic potential takes the
form of an adiabatic drift. At $B\sim B'_{\rm ad}$ the disorder
starts to dominate over the non-adiabatic effects, leading to a 
$B^{3/4}$-increase of the oscillation amplitude.
 }
\label{f8.2}
\end{figure}

\subsection{Low-field magnetoresistance}
\label{s8.3}

A distinct low-field magnetoresistivity was observed, along with the
commensurability oscillations, in the original experiment
\cite{weiss}, as well as in numerous later experiments on the
transport in a lateral superlattice. Specifically,  in low magnetic
fields $B$ a positive magnetoresistivity was found, followed by a
maximum in $\rho_{xx}(B)$. For not too strong modulation, the relevant
magnetic fields are much weaker than those where the Weiss
oscillations are observed, so that the two effects can be easily
separated. Soon after the first experimental observation it was
understood \cite{beton90} (see also 
\cite{menne98}) that the low-field magnetoresistivity is
related to the existence of open (channeled) orbits in the magnetic
fields $B<B_c=(\eta c/2e)qmv_F$. It is worth mentioning that this
effect, which is not found within the  $\eta$-expansion used in
Refs.~\cite{beenakker,mw98}, has its counterpart in the context
of the sound absorption in metals in the presence of a magnetic
field. There, the trapping of electrons in channeled  orbits by a
sound wave leads to non-linearity of the acoustic response of an
electron gas, as was observed experimentally \cite{fil} and analyzed
theoretically \cite{galperin}. 

A quantitative analytical description of the problem in the presence of
disorder was worked out in \cite{mtw2}. It was found that for a
suffuciently strong modulation, $\eta^{3/2}ql\gg 1$, the
contribution of channeled orbits to resistivity has the form 
\be
\label{e2.36}
\Delta\rho_{xx}^{\rm ch} / \rho_0 =
(\sqrt{2} / \pi^2)\eta^{7/2}(ql)^2 F_{\rm ch}(\beta)\ ,
\ee
where $\beta=B/B_c$ and $ F_{\rm ch}(\beta)$ is a parameterless
function shown in the left panel of Fig.~\ref{f8.3}. This induces a low-field
magnetoresistance that scales as
$\eta^{7/2}$ with the modulation strength.
It was further shown in \cite{mtw2} that the contribution of
non-channeled orbits is also modified at $B\lesssim B_c$, 
\be
\label{e2.18}
\Delta\rho_{xx}^{\rm nc} / \rho_0 = 2\pi\omega_c\tau 
\langle v_d^2\rangle / v_F^2= (\eta^2 / 2\pi) ql F_{\rm nc}(B/B_c)\ ,
\ee
where the dimensionless function $F_{\rm nc}(\beta)$ is shown in the
middle panel of Fig.~\ref{f8.3}.  
In the right panel of Fig.~\ref{f8.3} the theoretical
results are compared with experimental data of
Ref.~\protect\cite{albrecht}.

\begin{figure}[htb]
\includegraphics[width=0.3\textwidth]{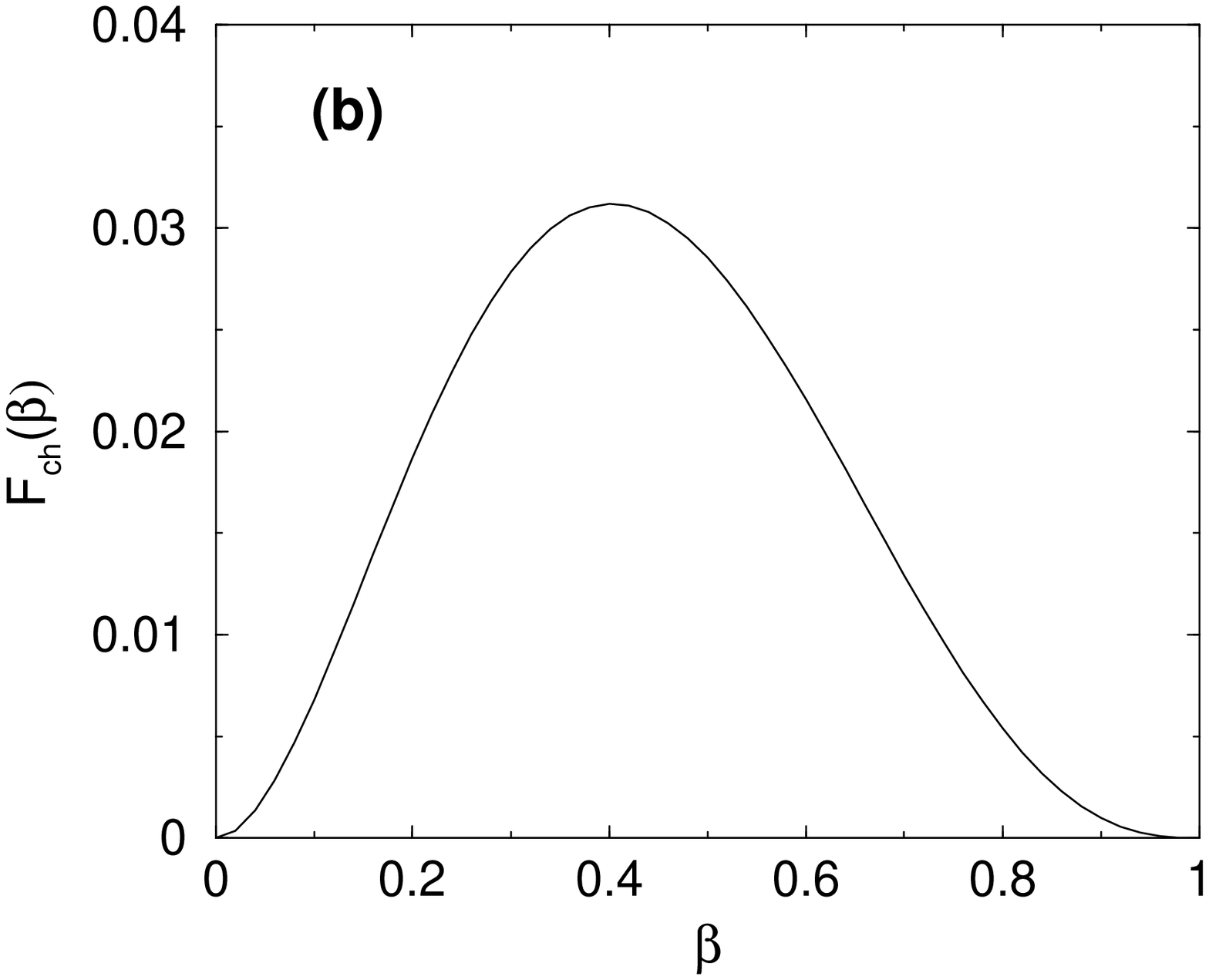} \hskip0.5cm
\includegraphics[width=0.3\textwidth]{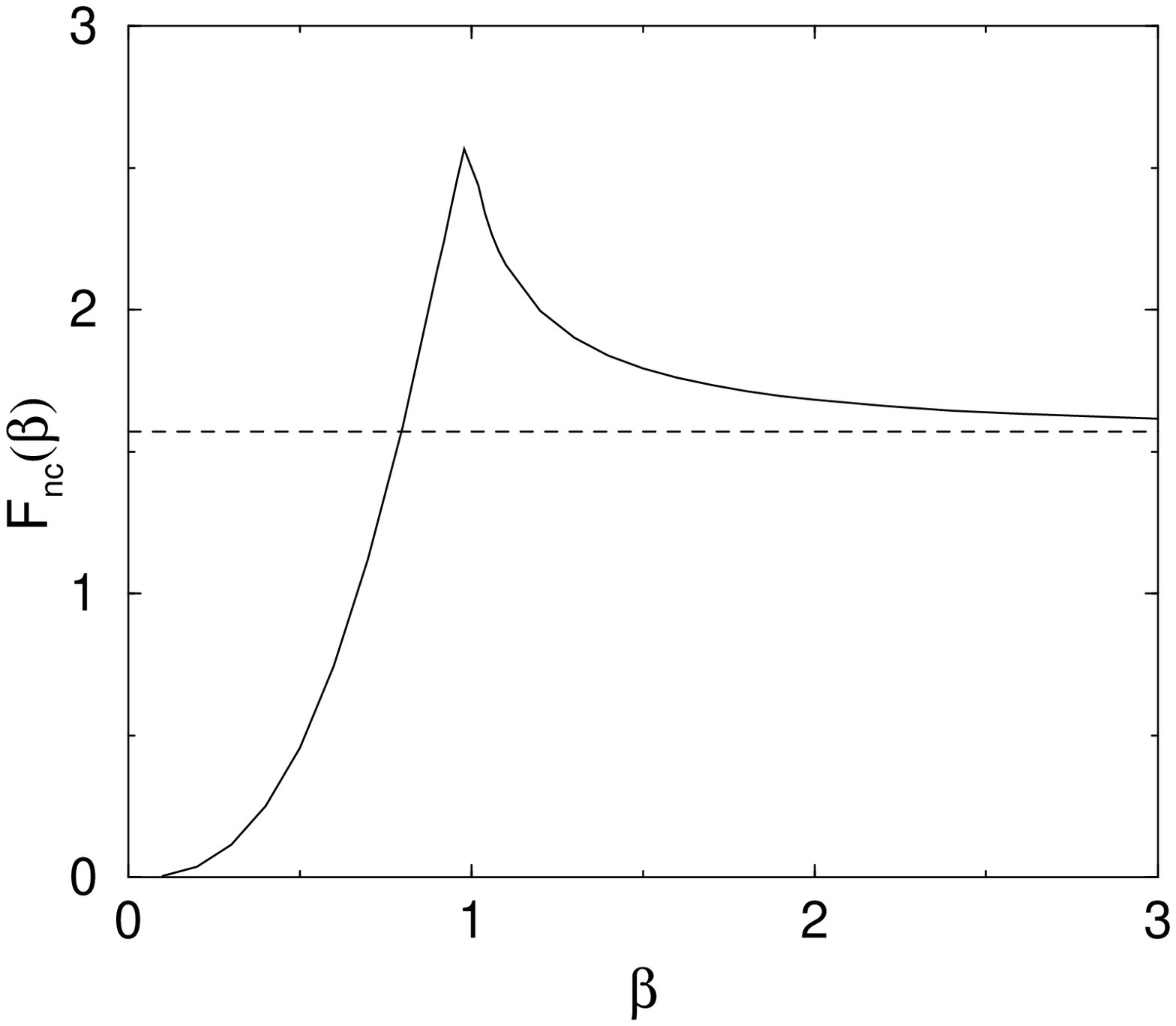} \hskip0.5cm
\includegraphics[width=0.3\textwidth]{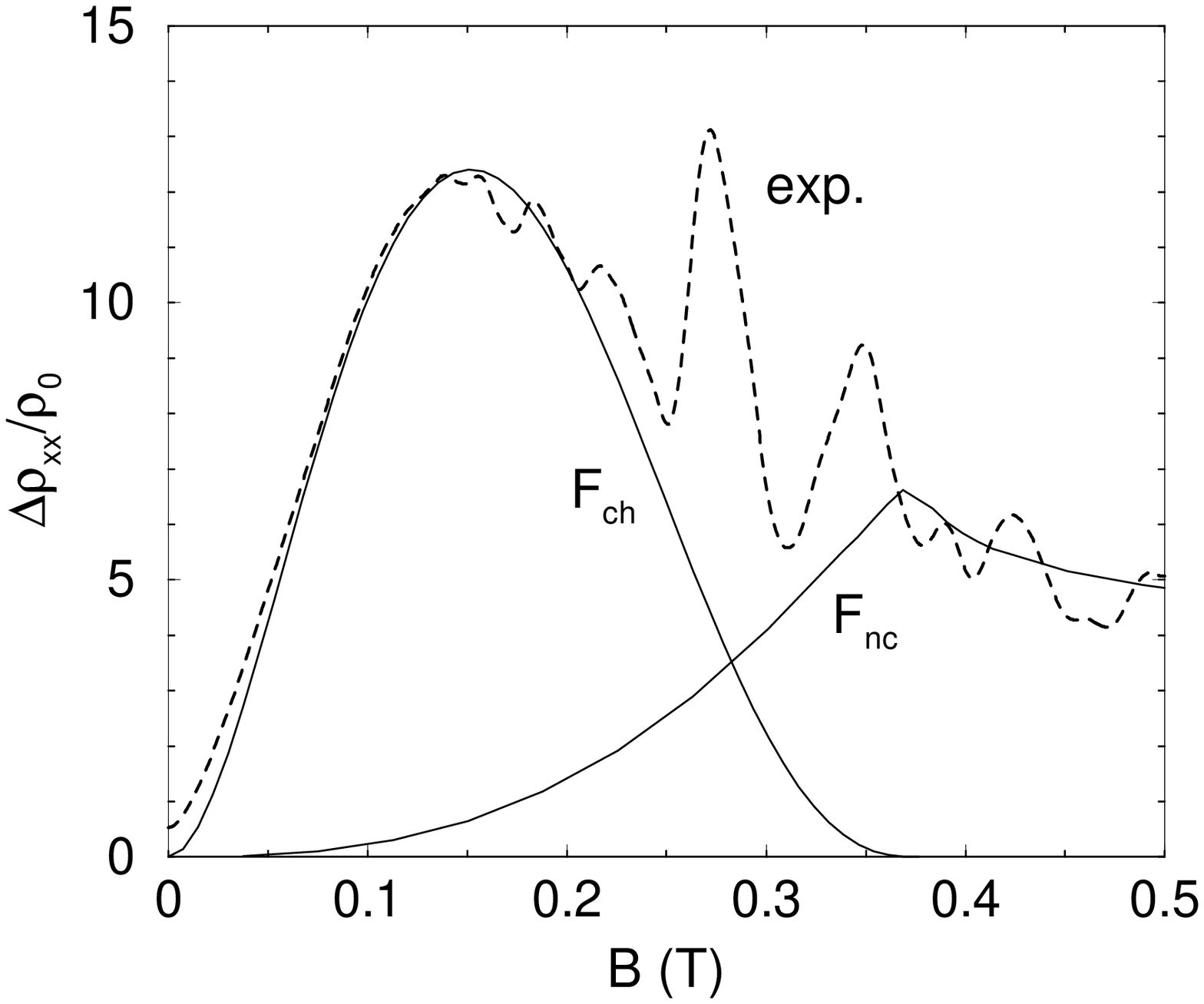}
\caption{
{\it Left panel:}
Function $F_{\rm ch}(\beta)$ describing magnetic field
dependence of the contribution of channeled orbits to the resistivity.
{\it Middle panel:} Function $F_{\rm nc}(\beta)$ characterizing
  the magnetic 
field dependence of the contribution of non-channeled orbits to
resistivity. The dashed line indicates the asymptotic value 
$F_{\rm nc}(\beta\gg 1)=\pi/2$.
{\it Right panel:}
Experimental data for the low-field magnetoresistivity from
Ref.~\protect\cite{albrecht} (dashed curve)
 compared to the theoretical results for the
contributions of channeled [Eq.(\ref{e2.18})] and non-channeled
[Eq.(\ref{e2.36})] orbits. The characteristic field $B_c$ and the
modulation amplitude found from the fit are $B_c\simeq 0.37\:{\rm T}$ and
$\eta\simeq 0.16$. 
}
\label{f8.3}
\end{figure}

\section{Interaction-induced magnetoresistance in non-quantizing fields}
\label{sec4}

As discussed in Sec. \ref{sec2},  
the longitudinal resistivity of an isotropic degenerate system is  
$B$--independent within the Drude-Boltzmann theory, $\rho_{xx}(B)=\rho_0=(e^2\nu v_F^2 \tau_{\text{tr}})^{-1}$,  
where $\nu_0$ is the density of states per spin.
There are several distinct sources of a non-trivial MR,  
which reflect the rich physics of  
2D systems. First, quasiclassical  
memory effects may lead to a MR(see Sec.\ref{sec2}), which shows no 
$T$-dependence at low temperatures. 
Second, weak localization~\cite{altshuler} induces a negative quantum MR 
restricted to the range of weak magnetic fields.  
Another quantum-mechanical source of the MR is electron--electron interaction. 
This type of the MR is the subject of the present Section.

It was discovered by Altshuler and Aronov \cite{altshuler}  
that the Coulomb  
interaction enhanced by the diffusive motion of electrons  
gives rise to a quantum correction to conductivity,  
which has in 2D the form  
$
\delta\sigma_{xx} \simeq (e^2/2\pi^2)  
\ln T\tau_{\text{tr}}  
$
(it is assumed here for simplicity that $\kappa \ll k_F$, 
where $\kappa=4\pi e^2\nu_0$ is  
the inverse screening length; we also set $k_B=\hbar=1$).  
The condition $T\tau_{\text{tr}}\ll 1$ under which this result  
is derived \cite{altshuler} implies that electrons move diffusively  
on the time scale $1/T$  
and is termed the ``diffusive regime''.  
Subsequent work \cite{SenGir} showed that in a strong magnetic field
this correction (in combination with  
$\delta\sigma_{xy}=0$) gives rise to a parabolic  
interaction--induced quantum  
MR,  
\begin{equation} 
\delta\rho_{xx}(B)/\rho_0 \simeq   
[(\omega_c\tau_{\text{tr}})^2-1] (\pi k_F l)^{-1} \ln T\tau_{\text{tr}},  \; \qquad  T\tau_{\text{tr}} \ll 1,  
\label{MRAA}  
 \end{equation}  
where $\omega_c=eB/mc$ is the cyclotron frequency and $l=v_F\tau_{\text{tr}}$ the   
transport mean free path. 
  
The effect of interaction on the conductivity  
in the ``ballistic regime'' $T> 1/\tau_{\text{tr}}$ has attracted a  
great deal of interest in a context of 2D systems showing a seemingly   
metallic behavior, $d\rho/dT>0$~\cite{kra,pud}.  
Zala, Narozhny, and Aleiner \cite{ZNA} developed  
a systematic theory of the interaction corrections  
valid for arbitrary $T\tau_{\text{tr}}$.  
In the ballistic range of temperatures, this theory 
predicts a linear-in-$T$ correction to   
conductivity $\sigma_{xx}$ and a $1/T$ correction to the Hall coefficient  
$\rho_{xy}/B$ at $B \to 0$, and describes the MR in a {\it parallel} field. 
The consideration of \cite{ZNA} is restricted, however,  
to {\it classically weak} transverse fields,   
$\omega_c\tau_{\text{tr}} \ll 1$, and to the {\it white-noise} disorder.  

In this Section,
we present a general theory of the interaction--induced corrections
to the conductivity tensor of 2D  
electrons valid for arbitrary $T$, $B$ and type of disorder \cite{GM}.   
A general expression for   
$\delta\sigma_{\alpha  \beta}$ is derived in terms of the ballistic propagator  
$D(\omega,{\bf q};{\bf n},{\bf n}')$ describing
the quasiclassical propagation of an electron in the phase space  
(${\bf n}$ is the unit vector characterizing the direction of 
velocity on the Fermi surface). 
The result for the exchange contribution reads  
  \begin{equation}  
\delta\sigma_{\alpha  \beta}=-2e^2 v_{\rm F}^2\nu_0 \int_{-\infty}^\infty  
\frac{d\omega}{2\pi}  
\frac{\partial}{\partial \omega}  
\left\{\omega\  {\rm coth}\frac{\omega}{2T}\right\} 
\int \frac{d^2{\bf q}}{(2\pi)^2}\  {\rm Im}  
\left[\ U(\omega,{\bf q})\  B_{\alpha  \beta}(\omega,{\bf q})\ \right],  
\label{sigma}  
\end{equation}  
where $U(\omega,{\bf q})$ is the interaction potential equal to  
a constant $U_0$ for point-like interaction and to  
$
U(\omega,{\bf q})=  
2\pi e^2\{q+\kappa[1+i\omega\langle D(\omega,q)\rangle]\}^{-1} 
$
for screened Coulomb interaction. 
The angular brackets $\langle \dots \rangle$ 
denote averaging over velocity directions ${\bf n},{\bf n}'$. 
The tensor $B_{\alpha  \beta}(\omega,{\bf q})$   
is given by  
\begin{eqnarray}
\label{Bwq_general}
B_{\alpha\beta}(\omega,{\bf q}) &=&  T_{\alpha\gamma} \pi\nu_0[ 
\langle D S_{\gamma\delta} D\rangle  -  
2\langle Dn_{\gamma} W n_{\delta}D\rangle] T_{\delta\beta} 
+ 
T_{\alpha\gamma}\left(\delta_{\gamma\delta}\langle D\rangle /2 
- \langle n_{\gamma} D n_{\delta} \rangle \right) T_{\delta\beta}
\nonumber \\
&-&2 T_{\alpha\gamma} \langle n_{\gamma} D n_{\beta} D \rangle  
- \langle D  n_{\alpha} D n_{\beta} D \rangle,
\end{eqnarray}
where
$ 
T_{\alpha  \beta}=2\left.\langle n_\alpha  
Dn_  \beta\rangle\right|_{q=0,\omega\to 0}= 
\sigma_{\alpha  \beta}/e^2v_F^2\nu_0,\ 
$
$S_{xx}=S_{yy}=W({\bf n},{\bf n'}), 
\quad S_{xy}=-S_{yx}=\omega_c/2\pi\nu_0$,
and $W({\bf n},{\bf n'})$ is the impurity scattering cross-section.
At $B\to 0$ one recovers the results for $\delta\sigma_{xx}$
and $\rho_{xy}$  
obtained in a different way in \cite{ZNA} for a white-noise disorder.  
Needless to say, in the diffusive limit,   
Eqs.~(\ref{sigma}), 
(\ref{Bwq_general}) reproduce (for arbitrary $B$ and disorder range)   
the logarithmic correction (\ref{MRAA}). 
 
The structure of Eqs.~(\ref{sigma}), 
(\ref{Bwq_general}) implies that the interaction correction is governed by 
returns of a particle to the original 
point in a time $t< T^{-1}\ll \tau_{\text{tr}}$. 
In a smooth random potential with 
a correlation length $d\gg k_F^{-1}$ the return probability  
is exponentially suppressed for $t\ll \tau_{\text{tr}}$. Therefore, the 
interaction correction in the ballistic regime
is exponentially small at $B=0$ for the case of 
smooth disorder. Moreover, the same argument applies to the case of
a non-zero $B$, as long as $\omega_c \ll T$. 
 
The situation changes qualitatively in a strong $B$, 
$\omega_c \gg T,\tau_{\text{tr}}^{-1}$: 
the particle experiences within the time $t\sim T^{-1}$ multiple 
cyclotron returns to the region close to the starting point. 
The MR is then determined by the correction to $\sigma_{xx}$.
For the Coulomb interaction, the exchange contribution to the
MR is given by 
\begin{equation}
\delta\rho_{xx}/\rho_0=
-(\omega_c\tau_{\text{tr}})^2 G_F(T\tau_{\text{tr}})/\pi k_F l.
\label{exchange}
\end{equation} 
For the point-like interaction a similar
result is obtained, with the replacement
$G_F(T\tau_{\text{tr}}) \to \nu_0 U_0  G_0(T\tau_{\text{tr}})$.
The functions $G_0(T\tau_{\text{tr}})$ and $G_F(T\tau_{\text{tr}})$, 
governing the $T$-dependence of the MR, are shown in Fig.~\ref{fig1}.
In the diffusive ($T\tau_{\text{tr}}\ll 1$) and ballistic ($T\tau_{\text{tr}}\gg 1$) limits 
they have the following asymptotics
\begin{equation}
\label{G0}
G_0(x)\simeq \left\{\begin{array}{ll}  
-\ln x + {\rm const}, & \ \ \ x\ll 1,\\  
c_0x^{-1/2}, & \ \ \ x \gg 1,
\end{array}\right. \quad
G_{\rm F}(x)\simeq \left\{\begin{array}{ll}  
\displaystyle
-\ln x + {\rm const}, & \ \ \ x\ll 1,\\  
\displaystyle
c_0x^{-1/2}/2, & \ \ \ x \gg 1,
\end{array}\right.
\end{equation}
where $c_0=3\zeta(3/2)/16\sqrt{\pi}\simeq 0.276$

\begin{figure}[htb]
\begin{minipage}[t]{.49\textwidth}
\includegraphics[width=0.49\textwidth]{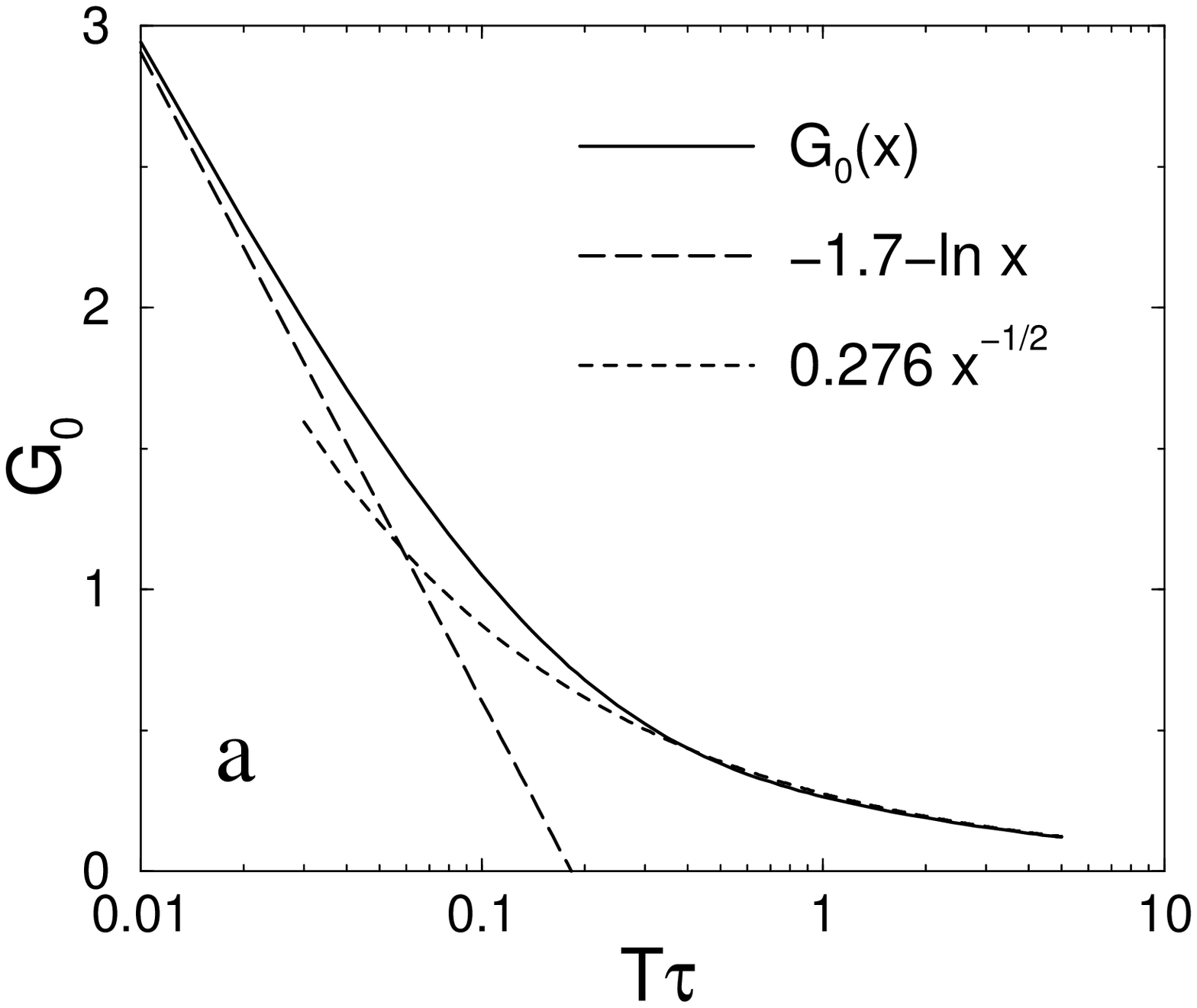}
\includegraphics[width=0.49\textwidth]{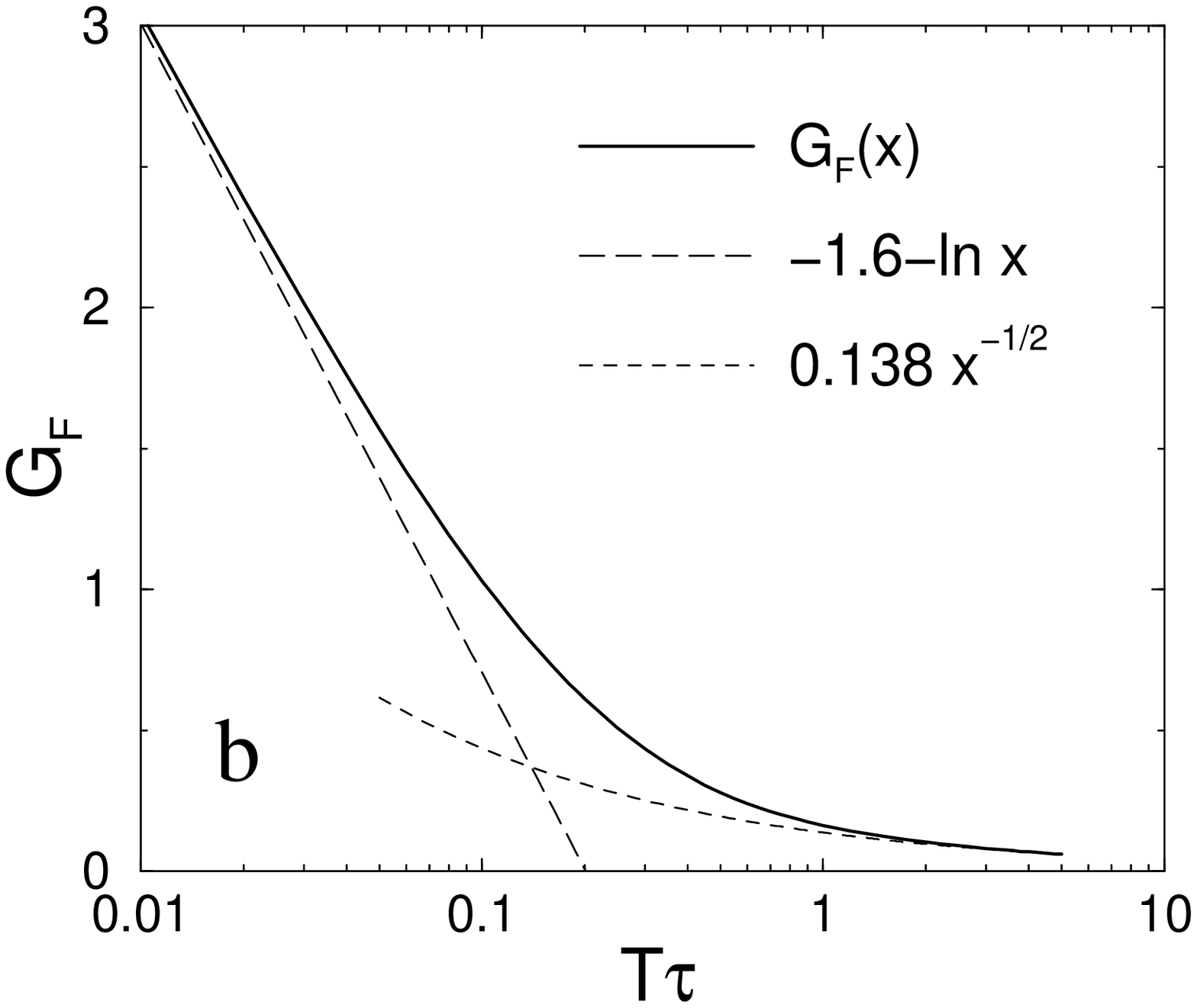}
\caption{Functions $G_0(T\tau_{\text{tr}})$ (a) and $G_{\rm F}(T\tau_{\text{tr}})$ (b)
determining the $T$-dependence of the exchange term for point-like and Coulomb
interaction, respectively, Eq.~(\ref{G0}).}
\label{fig1}
\end{minipage}
\hfill
\begin{minipage}[t]{.49\textwidth}
\includegraphics[width=0.49\textwidth]{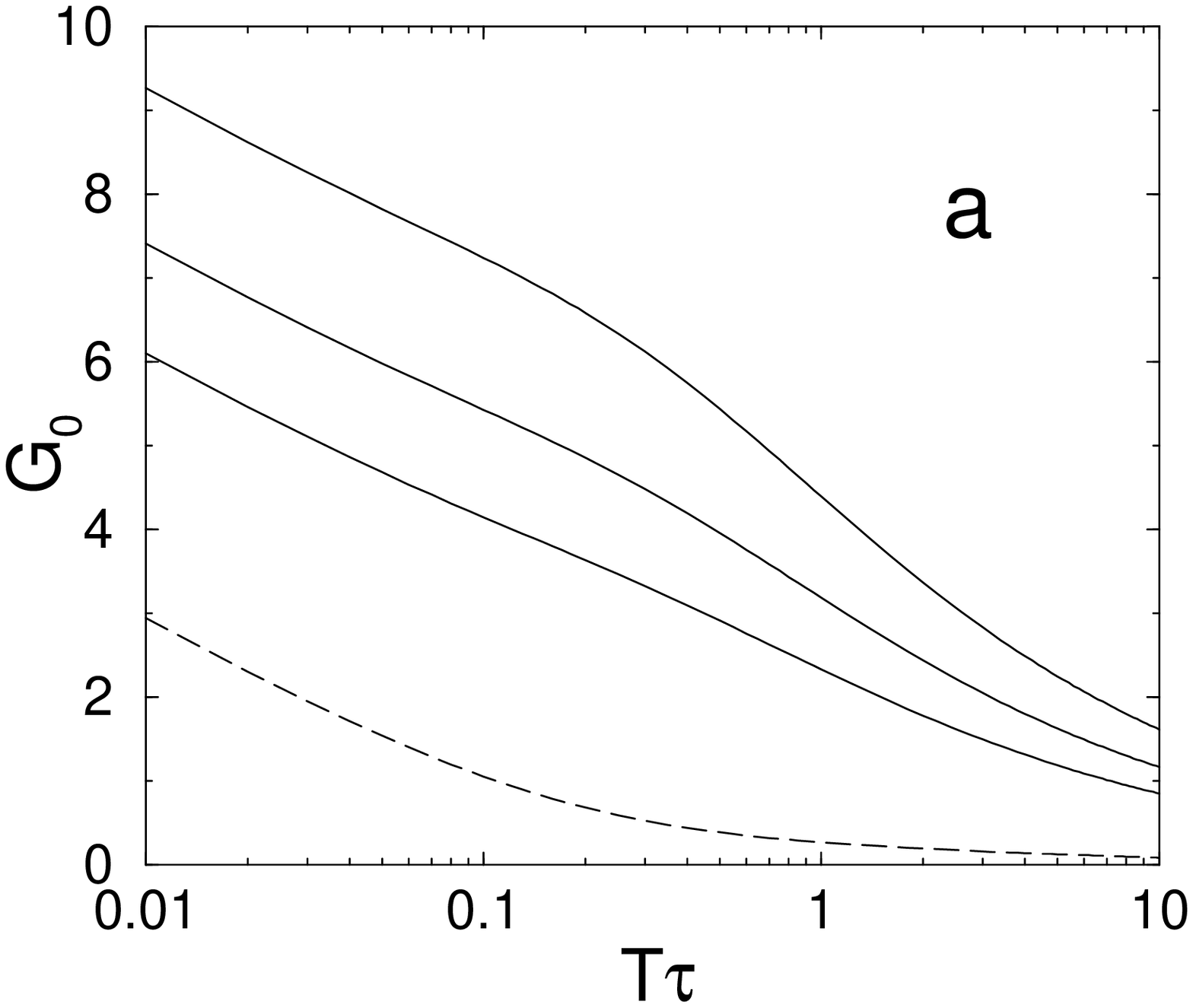} 
\includegraphics[width=0.49\textwidth]{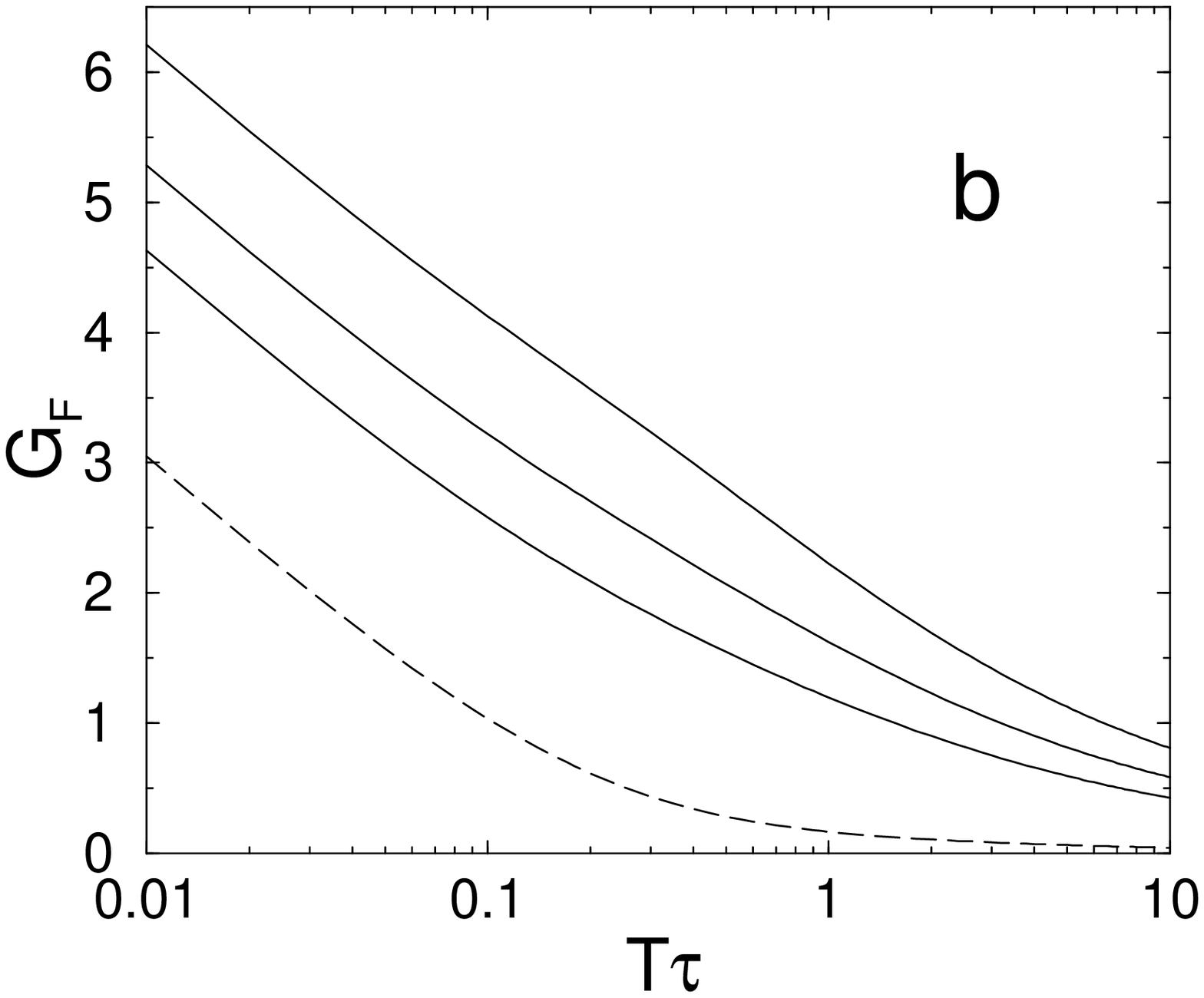} 
\caption{Functions $G_0^{\rm mix}(T\tau_{\text{tr}})$ (a) and  $G_F^{\rm mix}(T\tau_{\text{tr}})$ (b)
describing the $T$ dependence of the MR for point-like and Coulomb
 interaction, respectively, in the mixed-disorder model
 for different values of parameter $\gamma\equiv \tau_{\rm sm}/\tau_{\text{tr}}=20,\ 10,\ 5$
 (from top to bottom).  Dashed curves represent these functions for 
 purely smooth disorder ($\gamma=1$).}
\label{fig4}
\end{minipage}
\end{figure}

We turn now to the Hartree term, 
assuming $\kappa\ll k_F.$ The expression for its triplet part
is analogous to (\ref{sigma}) with the replacement of 
$U(\omega, {\bf q})$ by $-\frac{3}{2}U(0,2k_F\sin[(\phi-\phi')/2])$, where 
$\phi$ and $\phi'$ are the angles of the electron velocity.
As to the singlet part, it is renormalized by mixing with the exchange
term. The total Hartree contribution (see Fig.~2) reads 
\begin{equation}
\frac{\delta\rho^{\rm H}_{xx}(B)}{\rho_0}=
 frac{(\omega_c\tau_{\text{tr}})^2}{\pi^2 k_Fl}
\left\{\begin{array}{ll}  
y\ln y\left[{3\over 4}\ln(T\tau_{\text{tr}})+\ln y\right], \quad & \  T\tau_{\text{tr}}\ll 1,\\  
y\ln^2[y(T\tau_{\text{tr}})^{1/2}], \quad & \  1 \ll T\tau_{\text{tr}} \ll (k_F/\kappa)^{2}, \\
{\pi c_0 (T\tau_{\text{tr}})^{-1/2}}, \quad &  \  T\tau_{\text{tr}} \gg (k_F/\kappa)^{2}.
\end{array}\right. 
\end{equation}
If $\kappa/k_F$ is not small, 
the exchange contribution remains unchanged, while 
the Hartree term
is subject to strong Fermi-liquid renormalization 
\cite{altshuler,ZNA} and is determined by angular 
harmonics $F_m^{\sigma,\rho}$
of the Fermi-liquid interaction $F^{\sigma,\rho}(\theta).$ 
The theoretically predicted $T^{-1/2}$ dependence of the interaction-induced
MR has been observed in a high-mobility n-GaAs heterostructure
with the smooth disorder ~\cite{Sav}.

Calculation of the correction 
$\delta\rho_{xy}$ to the Hall resistivity requires evaluation of
both $\delta\sigma_{xx}$ and $\delta\sigma_{xy}$. 
The temperature dependence of $\delta\rho_{xy}$
in a strong $B$ is governed by 
$\delta\sigma_{xx}$ in the diffusive limit and
by $\delta\sigma_{xy}$ in the ballistic limit. 
For the point-like interaction, we find
\begin{equation}
{\delta\rho_{xy}/ \rho_{xy}}=
{\nu_0 U_0} G_0^{\rho_{xy}}(T\tau_{\text{tr}})/{\pi k_F l}, \qquad
G_0^{\rho_{xy}}(x)=\left\{\begin{array}{ll}  
-2 \ln x +{\rm const}, & \ \ \ x\ll 1,\\  
- 11 c_1 x^{1/2}, & \ \ \ x \gg 1,
\end{array}\right. \label{G0rhoxy}
\end{equation}
with $c_1=-\sqrt{\pi}\zeta(1/2)/4\simeq 0.647$.
An analogous consideration for the Coulomb interaction 
yields a similar result for the exchange correction
(Fig.~\ref{fig:3xy})
\begin{equation}
{\delta\rho^{\rm F}_{xy}/ \rho_{xy}}=
{G_{\rm F}^{\rho_{xy}}(T\tau_{\text{tr}})}/{\pi k_F l}, \qquad
G_{\rm F}^{\rho_{xy}}(x)
 = \left\{\begin{array}{ll}  
-2 \ln x +{\rm const}, & \ \ \ x\ll 1,\\  
\displaystyle 
- (11/2) c_1 x^{1/2}, & \ \ \ x \gg 1.
\end{array}\right. \label{rhoxy-exchange}
\end{equation}
\begin{SCfigure}[4][htb]
\includegraphics[width=.3\textwidth]{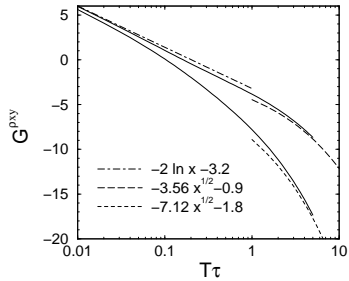}
\caption{Functions  $G_0^{\rho_{xy}}(T\tau_{\text{tr}})$ (lower curve)and
  $G_F^{\rho_{xy}}(T\tau_{\text{tr}})$ (upper curve) describing the temperature 
 dependence of the Hall resistivity for point-like and Coulomb interaction,
 respectively. Diffusive ($x \ll 1$) and ballistic ($x \gg 1$) 
 asymptotics are also shown.}
\label{fig:3xy}
\end{SCfigure}

Above the interaction correction for a system
with a small-angle scattering induced by smooth disorder with
correlation length $d\gg k_F^{-1}$ has been studied. This is a typical situation
for high-mobility GaAs structures with sufficiently large spacer $d$.
It is known, however, that with further increasing width of the spacer
the large-angle scattering on residual impurities and interface 
roughness becomes important and limits the mobility (see Sec.~\ref{s2.2}).
Furthermore, in Si-based structures the transport relaxation
rate is usually governed by scattering on short-range impurities.
This suggests considering the two-component model of disorder:
white-noise random potential with a mean free time $\tau_{\rm wn}$
and a smooth random potential with a transport relaxation time
$\tau_{\rm sm}$.
It is assumed that while the transport relaxation rate 
$\tau_{\text{tr}}^{-1}=\tau^{-1}_{\rm wn}+\tau^{-1}_{\rm sm}$ is 
governed by short-range disorder, $\tau_{\rm wn}\ll \tau_{\rm sm}$,
the damping of SdHO is dominated by smooth random potential.
This allows one to consider the range of classically strong magnetic fields,
$\omega_c\tau_{\rm wn}\gg 1,$ neglecting at the same time Landau
quantization.

The presence of short-range scatterers enhances the MR
as compared to the case of smooth disorder.
For the point-like interaction the MR reads
\[
{\delta\rho_{xx}(B)\over\rho_0}= 
-{(\omega_c\tau_{\text{tr}})^2 \nu_0 U_0\over \pi k_Fl}
G_{0}^{\rm mix}\left(T\tau_{\text{tr}},{\tau_{\rm sm}\over \tau_{\text{tr}}}\right), \quad
G^{\rm mix}_0(x,\gamma) 
=\left\{\begin{array}{ll}  
-\ln x + (2\gamma)^{1/2}, &  \ x\ll 1,\\  
4 c_0 \gamma^{1/2} x^{-1/2}, &  \ x \gg 1.
\end{array}\right.
\]
In the case of Coulomb interaction, 
the exchange contribution is given by
\[
{\delta\rho^{\rm F, mix}_{xx}(B)\over \rho_0}=
-{(\omega_c\tau_{\text{tr}})^2\over  \pi k_Fl}
G_{\rm F}^{\rm mix}\left(T\tau_{\text{tr}},{\tau_{\rm sm}\over\tau_{\text{tr}}}\right), \quad 
G_{\rm F}^{\rm mix}(x,\gamma) 
=\left\{\begin{array}{ll}  \displaystyle
-\ln x + (\gamma/2)^{1/2}, & \  x\ll 1,\\[0.2cm]  
2 c_0 \gamma^{1/2} x^{-1/2}, & \ x \gg 1.
\end{array}\right.
\]
These results (illustrated in Fig.~\ref{fig4}) are in a good agreement with experimental
data of Ref.~\cite{kvon} for mixed disorder in a Si/SiGe heterostructure. 
Finally, the ballistic contribution
to $\rho_{xy}$ is also enhanced by a factor 
$\sim(\tau_{\rm sm}/\tau_{\text{tr}})^{1/2}$ in the mixed disorder model.

The formalism can be further applied to anisotropic systems.  
The interaction-induced correction mixes the components $\rho_{xx}$ and
$\rho_{yy}$ of the resistivity tensor. This result is of special
interest in the case of systems subject to a one-dimensional periodic
modulation (lateral superlattice, Sec.~\ref{s8}; wave vector 
${\bf k}\parallel {\bf e}_x$). Specifically, it has been shown that
the interaction induces novel oscillations in $\rho_{yy}$, which are
in phase with quasiclassical commensurability (Weiss) oscillations in
$\rho_{xx}$.

\section{Influence of Landau quantization on magnetotransport}
\label{sec3}
In this section, we address magnetooscillations in the dissipative dc
and ac conductivity of a 2DEG governed by the Landau quantization.
Despite these effects, to the first place, Shubnikov-de Haas
oscillations (SdHO), are well established experimentally, the
theoretical description until recently was only available \cite{ando} for fully
separated Landau levels (LLs) with point-like scatterers \cite{footAndo}. A
systematic approach to the problem was developed in Ref.~\cite{dmitriev03}.
The results of this work, valid also for overlapping LLs, and for experimentally relevant case of smooth disorder, with
the correlation length $d\gg k_F^{-1}$, are reviewed below.

Within the quasiclassical Boltzmann theory, the dissipative
{\it ac} conductivity $\sigma_\omega=\sigma_+(\omega)+\sigma_-(\omega)$
of a non-interacting 2DEG is given by the Drude
formula (we neglect spin for simplicity), 
\be
\label{drude}
\sigma^{D}_{\pm}(\omega) = (1/4){e^2\nu_0 v_F^2\tau_{\rm tr}/
  [1+(\omega_c\pm\omega)^2\tau_{\rm tr}^2]}~, \ee 
where $\nu_0=m/2\pi$ and $\tau_{\rm tr}$ are the density of states (DOS) and
the transport relaxation time at $B=0$, $\omega_c=eB/mc$ the
cyclotron frequency, and $m$ is the electron effective mass.
We consider a 2DEG subjected to quantizing magnetic field $B$ and a random potential
$U({\bf r})$ characterized by a correlation function $\langle
U(\br)U(\br')\rangle = W(|\br-\br'|)$. The total and
the transport relaxation rates at $B=0$ are 
\bea\nonumber
\label{rates}
\left.
\begin{array}{l}
\tau_{\rm q}^{-1} \\ 
\tau_{\rm tr}^{-1}
\end{array}
\right\}
= 2\pi\nu_0\int{d\phi\over 2\pi} \,\tilde{W}(2k_F\sin{\phi\over 2})\times
\left\{
\begin{array}{l}
1\\ 
(1-\cos\phi)
\end{array}
\right. , \eea 
where $\tilde{W}(\bq)$ is the Fourier transform of $W(\br)$.
While we are mainly interested in the experimentally relevant case of smooth
disorder, $d\gg k_F^{-1}$, with $\tau_{\rm tr}/\tau_{\rm q}\sim (k_F
d)^2\gg 1$, our results are valid for arbitrary $d$ (i.e., including
short-range disorder with $\tau_{\rm tr}/\tau_{\rm q}\sim 1$).
The conductivity is given by the Kubo formula
\be \sigma_\omega=-\,({e^2}/{4\pi
  V\omega})\textstyle{\int}\!
{d\ve}\,(f_\ve-f_{\ve+\omega}){\rm Tr}\:\overline{ \hat{v}_x(G^A_{\ve+\omega}-G^R_{\ve+\omega})
  \hat{v}_x(G^A_{\ve}-G^R_{\ve})}~,
\label{sigma1}
\ee where $f_\ve$ is the Fermi distribution, $G^{R,A}$ are the
retarded and advanced Green functions, the bar denotes impurity
averaging, and $V$ is the system area. At high LLs,
$\ve_F\gg \omega,\omega_c$, disorder can be treated within the self-consistent
Born approximation (SCBA) \cite{ando} provided the
disorder correlation length satisfies $d\ll l_B$ and $d\ll
v_F\tau_{\rm q}$, where $l_B=(c/eB)^{1/2}$ is the magnetic length
\cite{raikh}. The SCBA equations for the Green function in the LL
representation, $G_n^R = (G_n^A)^*$, read
\cite{ando,raikh}, \be
G^{R}_n(\ve)=(\ve-\ve_n-\Sigma_\ve)^{-1},\qquad
\Sigma_\ve=(\omega_c/2\pi\tau_{q,0}){\sum}_n G^{R}_n(\ve)~,
\label{born}
\ee where $\ve_n=(n+{{1}\over{2}})\omega_c$ is the n-th LL energy
(Fig.~\ref{fig:SCBA}a).  The  conductivity
 (\ref{sigma1})  is  given  by  an electronic bubble  with a vertex
 correction,  i.e., by a  sum of ladder diagrams,  
Fig.~\ref{fig:SCBA}b,c.  In the
  case of white-noise  disorder, $\tau_{\rm q}=\tau_{\rm tr}$,
  the vertex correction is zero, and  it suffices to
  evaluate  the  bare bubble
\be\nonumber
\!\!\!\sigma_\pm^b(\omega)\!=\!\frac{e^2 v_F^2\nu_0}{4}\!\!\int\!
{{d\ve}\over{\omega}}\,(f_\ve-f_{\ve+\omega})\,{\rm Re} (\Pi_\pm^{RA}-\Pi_\pm^{RR}),
  \qquad 
  \Pi_\pm^{RR(RA)}\!=\!{\omega_c\over 2\pi}
\sum_n G^R_{n\pm 1}(\ve+\omega)G^{R(A)}_n(\ve).
\ee
For the case of smooth disorder we have to take into account the
vertex correction (Fig.~1c) while averaging in Eq.~(\ref{sigma1}).
This is a non-trivial task since the disorder mixes strongly the LLs,
thus seriously complicating a direct calculation in the LL
representation. The result, however, acquires a remarkably simple and physically transparent form:
provided the above SCBA conditions are fullfilled, the inclusion of the vertex correction results in a
replacement of $\Pi_\pm^{RR(RA)}$ above by
$
\Pi_{\pm,{\rm tr}}^{RR(RA)}\equiv\left[\left(\Pi_\pm^{RR(RA)}\right)^{-1}
-(\tau_{\rm q}^{-1}-\tau_{\rm tr}^{-1})\right]^{-1}.
\label{PiTr}
$ 
It follows that $\sigma_\omega$ has a Drude-type
structure with the DOS, 
$\nu(\ve)=-\pi^{-1}\nu_0 \omega_c {\rm Im}\sum_n G^R_n(\ve),$
and the transport time renormalized due to Landau
quantization:
\be\label{Qdrude}
\sigma_\pm(\omega)=\frac{e^2 v_F^2}{4\omega}\!\int\!
\frac{d\ve\,(f_\ve-f_{\ve+\omega})\,\nu(\ve)\,\tau_{\rm    
    tr,B}^{-1}(\ve+\omega)}{[\tau^{-2}_{\rm tr,B}(\ve)+\tau^{-2}_{\rm
        tr,B}(\ve+\omega)]/2+(\omega\pm\omega_c)^2}~,\qquad \tau_{\rm
      tr,B}(\ve)\equiv\tau_{\rm tr}\frac{\nu_0}{\nu(\ve)}~.  \ee 
Formula (\ref{Qdrude}) is the main result of this section. 
Let us emphasizethat the single-particle time $\tau_{\rm q}$ enters Eq.~(\ref{Qdrude}) only
through the DOS; everywhere else it has been
replaced by the transport time $\tau_{\rm tr}$ due to the vertex correction. 
In the following, we analyze Eq.~(\ref{Qdrude}) in several important limiting cases.

\begin{figure}[htb]
\begin{minipage}[t]{.5\textwidth}
\includegraphics[width=\textwidth]{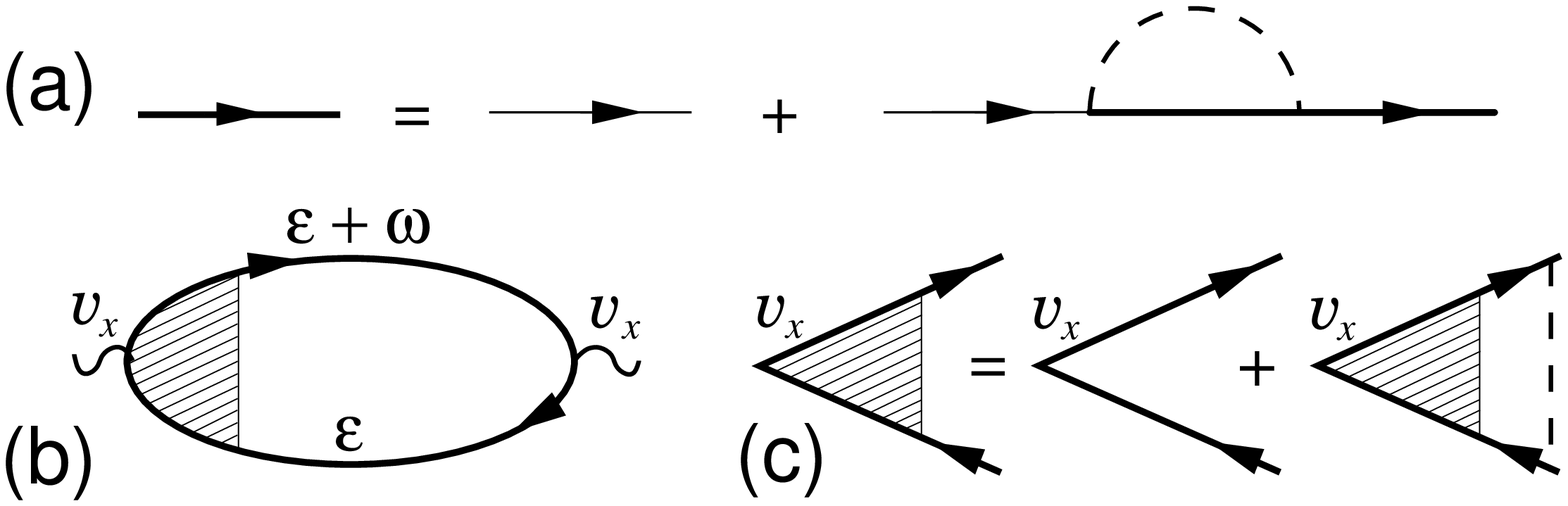}
\caption{(a) SCBA equation for the Green function;
(b) dynamical conductivity with vertex correction (c).}
\label{fig:SCBA}
\end{minipage}
\hfill
\begin{minipage}[t]{.4\textwidth}
\includegraphics[width=\textwidth]{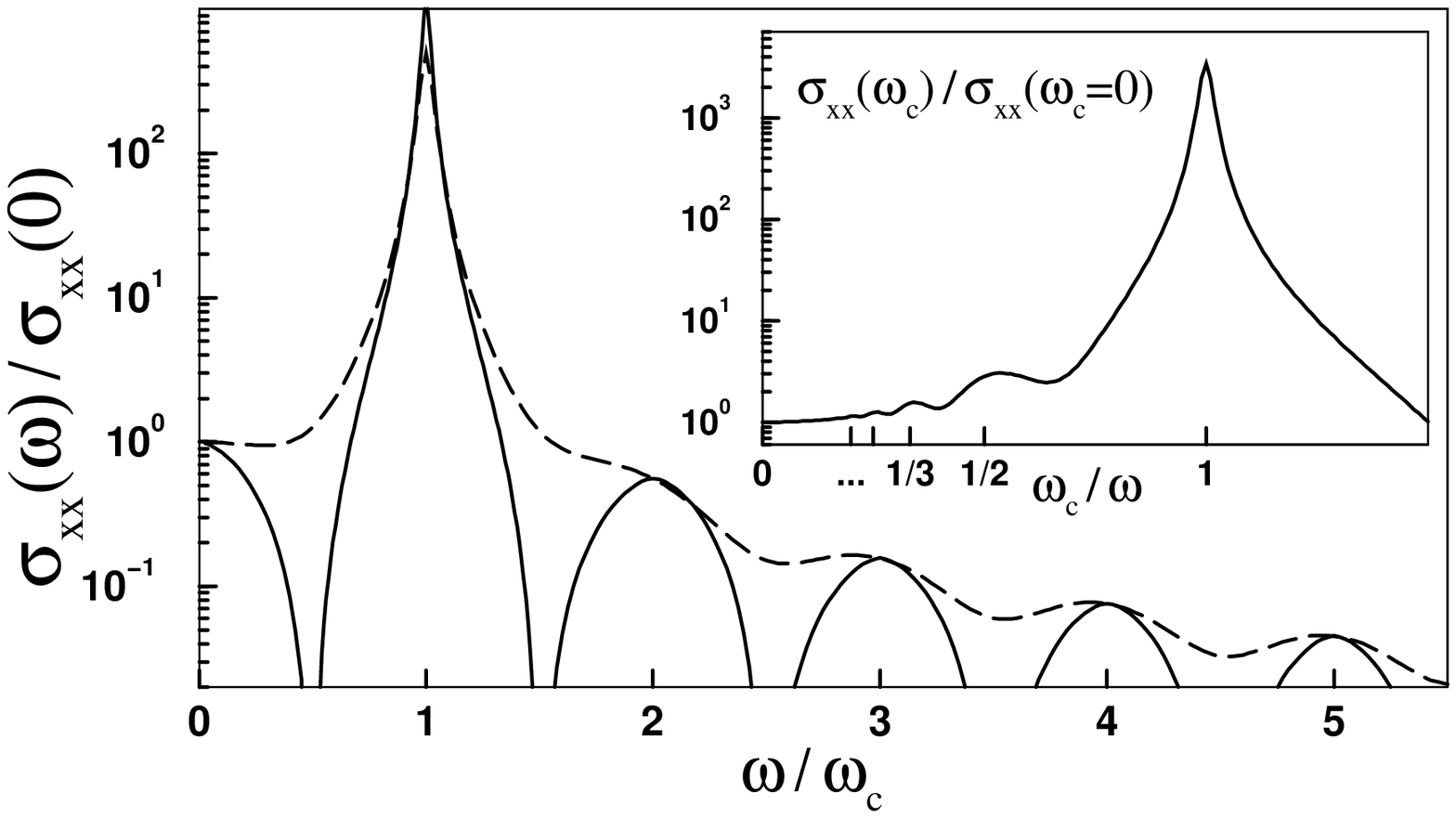}
\caption{Magnetooscillations of the dynamical conductivity for
  $\tau_{\rm tr}/\tau_{\rm q}=10$. Solid line: separated   LLs,
  $\omega_c\tau_{\rm q}/\pi=3.25$; dashed line: overlapping LLs,
  $\omega_c\tau_{\rm q}/\pi=1$. Inset: $\sigma_{xx}$ for fixed
  $\omega\tau_{\rm q}/2\pi=1$ as a function of $\omega_c$.}
\label{fig:qosc}
\end{minipage}
\end{figure}
%
 
In the regime of strongly overlapping LLs, $\omega_c\tau_{\rm q}\ll
1$, the solution to SCBA equations (\ref{born}) is most easily
obtained using the Poisson formula, $\sum_n F_n=\sum_k\int dx F(x)\exp
(2\pi i k x)$.  The $k=0$ term yields the $B=0$ result,
while the $k=\pm 1$ contributions provide the leading oscillatory correction
to the DOS,
\be \nu(\ve)=\nu_0[1-2\delta\cos(2\pi\ve/\omega_c)+{\cal O}(\delta^2)]~,
\qquad\delta=\exp(-\pi/\omega_c \tau_{\rm q})\ll 1~.
\label{nuOvLLs}
\ee
To first order in
$\delta$, Eq.~(\ref{Qdrude}) produces the following result: 
\be\label{corr1}
\frac{\sigma^{(1)}_\pm(\omega)}{\sigma^{D}_\pm(\omega)}=
  1 - 2\delta~{\cal F}\left({2\pi^2 T\over\omega_c}\right)\cos \frac{2\pi\ve_F}{\omega_c}
\left[\frac{2\alpha_\pm^2}{\alpha_\pm^2+1} \frac{\sin(
      2\pi\omega/\omega_c)}{2\pi\omega/\omega_c}+
    \frac{3\alpha_\pm^2+1}{\alpha_\pm^2+1}
    \frac{\sin^2(\pi\omega/\omega_c)}{\alpha_\pm\pi\omega/
      \omega_c}\right]~, 
\ee
where $\alpha_\pm\equiv\tau_{\rm tr}(\omega\pm\omega_c)$, and the
Dingle factor ${\cal F}(X)=X/\sinh X$ describes the $T$--damping of
the SdHO.  In the {\it dc} limit $\omega\to0$, this result confirms
the form of SdHO in smooth disorder
conjectured in \cite{col}.  If $T$ is higher than the Dingle
temperature $T_D\equiv 1/2\pi\tau_{\rm q}$, the temperature smearing
becomes the dominant damping factor. In high-mobility 2DEG the Dingle
temperature is as low as $T_D\sim 100\:{\rm mK}$, so that for
characteristic measurement temperatures $T\sim 1\:{\rm K}$ the
first-order correction (\ref{corr1}) will be completely suppressed.
However, there exists a correction of order $\delta^2$, oscillatory in
$\omega/\omega_c$, which is not affected by the temperature. To obtain
it, there is no need to calculate $\nu(\ve)$ to second order, since
the corresponding terms oscillate with $\ve$, and doesn't survive the
high--$T$ limit.  The leading quantum correction at $T\gg T_D$ results
from the averaging Eq.~(\ref{Qdrude}) over fast energy oscillations of
the first-order $\nu(\ve)$, Eq.~(\ref{nuOvLLs}), which gives
\be\label{corr2}  \sigma^{(2)}_\pm(\omega) =
\sigma^{D}_\pm(\omega)\left\{1+2\delta^2
  \left[\frac{\alpha_\pm^2(\alpha_\pm^2-3)}
    {(\alpha_\pm^2+1)^2}\cos\frac{2\pi\omega}{\omega_c} +
    \frac{\alpha_\pm(3\alpha_\pm^2-1)}{(\alpha_\pm^2+1)^2}
    \sin\frac{2\pi\omega}{\omega_c}\right]\right\}~.  \ee

The regime which is most interesting theoretically and relevant experimentally
is that of long-range disorder, $\tau_{\rm tr}/\tau_{\rm q}\gg 1$, and a
classically strong magnetic field, $\omega_c,\omega\gg \tau_{\rm tr}^{-1}$.
In this situation Eq.~(\ref{Qdrude}) reads
\be\label{Qdrude1}
\sigma_\omega=\sigma_\omega^D\!\int\!d\ve
\frac{f_\ve-f_{\ve+\omega}}{\omega\,\nu_0^2}\,\nu(\ve)\,\nu(\ve+\omega)~,
\qquad\sigma^{\rm D}_\omega=\sum_{\pm}\,\frac{e^2\nu_0 v_{\rm F}^2}
{4 \tau_{\rm tr}(\omega\pm\omega_c)^2}~,  \ee 
or, in {\it dc} limit,
\be\label{Qdrudedc}
\sigma_{\rm dc}=-\sigma_{\rm dc}^D\!\int\!d\ve [\nu^2(\ve)/\nu_0^2]
\partial_\ve f(\ve)~,\qquad\sigma^{\rm D}_{\rm dc}=e^2\nu_0 v_{\rm F}^2/
2 \tau_{\rm tr}\omega_c^2~,  
\ee 

In the limit of separated LLs, $\omega_c\tau_{\rm q}\gg 1$, the DOS is 
a sequence of semicircles of width $2\Gamma\ll\omega_c$, 
\be\label{nuSepLLs}
\nu(\ve)=\nu_0\tau_{\rm q}{\sum}_n {\rm Re}\sqrt{\Gamma^2-(\ve-\ve_n)^2}~,\qquad
\Gamma=\sqrt{2\omega_c/\pi\tau_{\rm q}}~.
\ee
In this case, $\sigma_\omega$ is non-zero only for $\omega$ in intervals
$[M\omega_c-2\Gamma,\:M\omega_c+2\Gamma]$ with an integer $M$.

Oscillations in $\sigma_\omega$ with $\omega/\omega_c$ for 2DEG with
smooth disorder at $\omega_c\tau_{\rm tr}\gg 1$ and $T\gg T_D$ are
illustrated in Fig.~\ref{fig:qosc}.  For overlapping LLs, oscillations
away from the cyclotron peak are described by simple formula
$\sigma_\omega/\sigma^{\rm
  D}_\omega=1+2\delta^2\cos(2\pi\omega/\omega_c)$.  For separated LLs,
at the center of the $M=1$ interval we find a CR peak of height
$\sigma_{xx}(\omega=\omega_c)= (e^2\nu_0 v_F^2/\pi\Gamma) \tau_{\rm
  tr,0}/\tau_{\rm q}$ and width $\sim \Gamma \tau_{\rm q}/\tau_{\rm
  tr,0}$. All other peaks ($M\ne 1$) are smaller by a factor
$\sim\tau_{\rm tr}/\tau_{\rm q}\sim(k_F d)^2\gg 1$, 
\be
\sigma_{xx}(\omega=M\omega_c)=(4e^2 \nu_0 v_F^2 \Gamma/3\pi\omega_c^2) 
(\tau_{\rm q}/\tau_{\rm tr})[(M^2+1)/(M^2-1)^2]~.  
    \ee

\section{Interaction effects in quantizing magnetic fields}
\label{sec6}
\subsection{Interaction effects on oscillations}

In this Section we discuss the interaction effects on magnetooscillations
(de Haas-van Alphen and Shubnikov -- de Haas oscillations), closely following
Ref.~\cite{yura}.
In addition to experimental motivation related to the 
apparent metal-insulator transition in 
two-dimensional systems, the corresponding theory complements 
the recently developed theory of
interaction effects in transport of 2D electrons in zero and
non-quantizing magnetic fields~\cite{ZNA,GM}.

The starting point is the
expression for the thermodynamic potential derived in the paper by Luttinger and Ward
\cite{LuttiWard60}
\be
  \Omega=-T{\text{Tr}}\ln(-G^{-1})-T{\text{Tr}} (G \Sigma)+\Omega',
  \label{LW}
\ee
where the trace implies summation over Landau levels $N$ and over 
fermionic Matsubara frequencies $\e_n=(2n+1)i\pi T$, 
$
\, G(i\e_n,N\omega_c)=[i\epsilon_{n}+\mu-(N+1/2)\omega_c-\Sigma(i\e_n,N\omega_c)]^{-1}
$
is the dressed Matsubara Green's function, 
and $\Sigma(i\e_n,N\omega_c)$ is a self-energy part of Green's
function which includes all the disorder and interaction effects.
The terms $-T{\text{Tr}} (G \Sigma)$ and $\Omega'$ in (\ref{LW})
are introduced to avoid double-counting of diagrams~\cite{LuttiWard60,AGD}.
The term $\Omega'$ denotes the sum of all so-called skeleton diagrams
with all bare Green's functions replaced by dressed Green's functions.

As shown in Ref.~\cite{luttinger}, the exponential decay of
magnetooscillations is described by the ${\text{Tr}}\ln$-term. The oscillatory parts
of the additional terms, which are introduced  to correct for
overcounting, cancel each other.
In order to obtain the correction to the thermodynamic potential
one needs to calculate the self-energy part of the Green's function.
It is worth noting that the inelastic contribution to the self-energy
$\propto [(\pi T)^2-\epsilon_n^2]{\text{sign}}\epsilon_n$ 
vanishes for $\epsilon_n=\pi T,$ and thus does not affect the damping of the
magnetooscillations $B(T)$ for $T\gg\omega_c,$
in agreement with Refs.~\cite{MMR,fowler}.

For overlapping Landau levels [Eq.~(\ref{nuOvLLs})],
the magnetooscillations are damped by disorder even at zero
temperature via
the standard Dingle factor $\delta = \exp(-\pi/\omega_c\tau_q)$.
Therefore we will consider only the first harmonics of the oscillations, $A_1$,
neglecting all higher harmonics whose damping is much stronger.
In what follows we concentrate on the case $T\gg\omega_c$.
Under this condition, the first harmonics of the oscillatory part of the
thermodynamic potential  
\be
\Omega_{\rm osc}
\simeq
2\nu_0 \left(\omega_c/2\pi\right)^2
A_1\cos(2\pi^2  n_e / e B),
\label{eq-omega-osc}
\ee
is given by
\begin{equation}
A_1=(4\pi^2 T/ \omega_c)\, \exp\left[-2\pi^2 T/\omega_c^*
- \pi/\omega_c^*\tau_q^*\right]\, \exp[B(T)],
\end{equation}
which is a standard FL Lifshitz-Kosevich expression
multiplied by the additional factor  with
\begin{equation}
\label{FT}
B(T)= -2\pi i \, Z \, \delta\Sigma(i\pi T,\xi_0)/\omega_c^*.
\end{equation}
Here $\omega_c^*=e B/ m^* $ is the Fermi-liquid
(FL) renormalized effective cyclotron frequency
in a pure system at zero $T$, which is related to the FL-renormalized 
effective mass $m^*$, $Z$ is the FL $Z$-factor (given by the residue of 
the Green's function), $\tau_q^*$ is the FL-renormalized scattering time, and 
$\delta\Sigma (i\e_n,\xi_0)$ is the self-energy part (taken
at the pole $\xi_0$ of the Green's function in
the presence of disorder)
describing the interplay of disorder and interaction.

\begin{SCfigure}[5]
  \includegraphics[width=0.35\columnwidth]{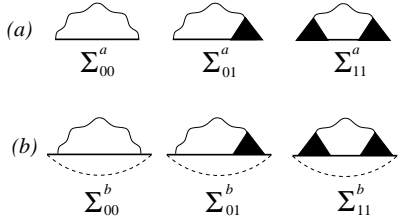}
      \caption{\label{f-sigma}  
Self energy diagrams in the first order in the effective interaction (wavy line).
Black triangles denote impurity ladders $\Gamma$ dressing interaction vertices, 
dashed line is a single-impurity line. Diagrams (b) represent the ``Hikami-box'' contribution to the self-energy which restores the gauge-invariance of the damping.}
\end{SCfigure}

Hereafter white-noise disorder with $\tau_{\text{tr}}=\tau_q\equiv \tau$
is considered.
Evaluating the sum of six digrams (Fig.~\ref{f-sigma}) for $\delta\Sigma$, 
one gets the following expression for the damping
exponent in the case of short-range interaction $U_0$:
\begin{equation}
 B(T)=- {\rm const}\, \nu_0 U_0 (\pi / \omega_c\tau_q)
 +
(\pi T/\omega_c)\,
(\nu_0 U_0/ \ve_F \tau) \ln(\ve_F/T).
\label{FT-short-result}
\end{equation}
The first term in Eq.~(\ref{FT-short-result}) describes the $T$-independent
FL-renormalization of $\tau_q$ due to vertex corrections and should be included in the
effective relaxation time $\tau_q^*$.
The second term represents the  $T$-dependent contribution to the
damping factor that we are interested in and  is analyzed below.

The above result (\ref{FT-short-result}) can be interpreted in terms
of corrections to the effective mass (or $\omega_c$) and the quantum elastic
scattering rate $\tau_q$
entering the standard Lifshitz-Kosevich formula. These corrections
come from the interplay of disorder and interaction, leading to
\begin{equation}
B(T)=-(2\pi^2 T/\omega_c)(\delta m/m)-(\pi/
  \omega_c\tau_q)
  \left[\delta m/m-
\delta\tau_q/\tau_q\right].
\label{FTviamtau}
\end{equation}
It is worth noting that the FL-renormalization does not affect the
product $\omega_cm=e B$.

Comparing (\ref{FT-short-result}) and  (\ref{FTviamtau}),
one can see that
the $T\ln T$ dependence of the damping factor
could in principle originate either from  the $\ln T$ correction
to the effective mass, or from the $T\ln T$-type correction to $\tau_q$.
This led the authors of Ref.~\cite{MMR} to the conclusion that
the nonlinear $T-$dependence of
the damping factor may be equivalently interpreted either as a
$T-$dependent renormalization of the
effective mass or as a $T-$dependent Dingle temperature.
It is clear, however, that these two possibilities correspond to different
physical processes. To identify the physical origin of the leading contribution to the damping
it is instructive to obtain $B(T)$ using the expression for
the self-energy analytically continued to real values
of energies $\epsilon_n\to -i\ve$ .

Having calculated $\text{Re}\Sigma$ and $\text{Im}\Sigma$ for real energies $\ve$, one can 
determine
$\delta m$ and $\delta \tau_q$. Indeed, the magnitude of the first harmonics of
the magnetooscillations of the thermodynamic density of states
is expressed through the real-$\ve$
self-energy $\delta\Sigma(\ve)$
as follows:
\bea
&&A_1(T)=-\int d\ve A_1(\ve,T) \partial_\ve f_T(\ve),\label{a1T}\\
&&A_1(\ve,T)=
\exp\left\{\frac{2\pi
    i}{\omega_c}[\ve-\text{Re}\delta\Sigma(\ve,\xi_0)]\right\}
\exp\left\{-\frac{\pi}{\omega_c\tau_q}
+\frac{2\pi}{\omega_c}\text{Im}\delta\Sigma(\ve,\xi_0)\right\}
\label{A1eT}\\
&&=\exp\left\{{2\pi i \ve \over\omega_c}\left[1+{\delta m(\ve,T)\over m}\right]\right\}
\exp\left\{-{\pi\over \omega_c\tau_q}\left[1+{\delta m(\ve,T)\over m}-
{\delta\tau_q(\ve,T)\over\tau_q}\right]\right\}.
\label{A1eTtm}
\eea
where $f_T(\ve)=[1+\exp(\ve/T)]^{-1}$ is the Fermi distribution function.
This allows one to express $\delta\tau_q(\ve, T)$ and
$\delta m(\ve,T)$
through $\text{Re} \Sigma(\ve)$ and $\text{Im} \Sigma(\ve)$ as follows:
\bea
\delta m(\ve,T)/ m&=& -\ve^{-1}\text{Re} \delta\Sigma(\ve,T)=
-  ({\nu_0 U_0}/{2 \pi \ve_F\tau}) \ln \{\ve_F/{\rm
     max}[|\ve|,T]\},
\label{deltamass}\\
{\delta\tau_q(\ve,T)/\tau_q}&=&  2 \tau_q \text{Im} \delta\Sigma(\ve,T)  +
{\delta m(\ve,T)/ m}\nonumber \\
&=&\nu_0 U_0 \frac{T}{ \ve_F} \ln\left[2 \cosh\left({\ve \over
        2T}\right)\right] - \frac{\nu_0 U_0}{2 \pi \ve_F\tau} \ln \frac{\ve_F}{{\rm
     max}[|\ve|,T]}.
\label{deltatautau}
\eea
It is clear from these results that
the leading term in $B(T)$ [proportional to $T \ln(\ve_F/T)$,
Eq.~(\ref{FT-short-result})] originates from the real part of the
self-energy, 
i.e. from renormalization of the effective mass,
which affects incommensurability
of the oscillations at different values of energy $\ve$.
The contribution of the imaginary part of the self-energy, which is
governed in the ballistic regime by the renormalization of the
scattering time, is smaller by a factor $\ln(\ve_F\tau)$.
The obtained result for the interaction-induced correction to the
quantum scattering time $\tau_q$, Eq.~(\ref{deltatautau}), agrees, up to a
factor ${1\over 2}$,  with the
correction to the transport time following from the calculation of
conductivity correction in the ballistic regime
in Ref.~\cite{ZNA}. 

In the case of Coulomb interaction, one should take into account the
dynamical screening of the interaction within the random phase approximation (RPA).
This leads
to different asymptotics of the self-energy in the diffusive
and ballistic
regimes, in contrast to the case of weak short-range
interaction.
The $T$-dependence of the leading
correction to the magnetooscillations
 damping factor due to the interaction in the singlet channel has
 the form (Fig.~\ref{Brho})
\be
B^\rho(T)=\frac{\pi}{\omega_{c}\tau_q}\frac{T}{\ve_F}  \times  \left\{
\begin{array}{ll} \displaystyle
({3 / 2})\ln({\ve_F}/{T}) - ({1/ 2}) \ln(4\pi \ve_F\tau) , & \quad  4\pi T\tau \ll 1,
\\[0.5cm]
\displaystyle
\ln({\ve_F}/{T}), & \quad 4\pi T\tau \gg 1.
\end{array}
\right.
\label{B-rho-T-lead-asympt}
\ee
Calculation of the corresponding triplet contribution leads to qualitatively similar asymptotics.
The leading term in the total correction to the damping factor in the ballistic regime,
realized in experiments on low-disorder
samples at realistic temperatures, takes the simple form
\be
B(T)=B^\rho(T)+B^\sigma(T)\simeq[1+{3F_0^\sigma}({1+F_0^\sigma})^{-1}]
\,({\pi T}/{\omega_{c}\tau_q \ve_F})\,\ln({\ve_F}/{T}).
\label{B-tot-T-lead}
\ee
As discussed above, this result arises due to the correction to the effective mass.
\begin{figure}[ht]
 \includegraphics[width=0.35\columnwidth]{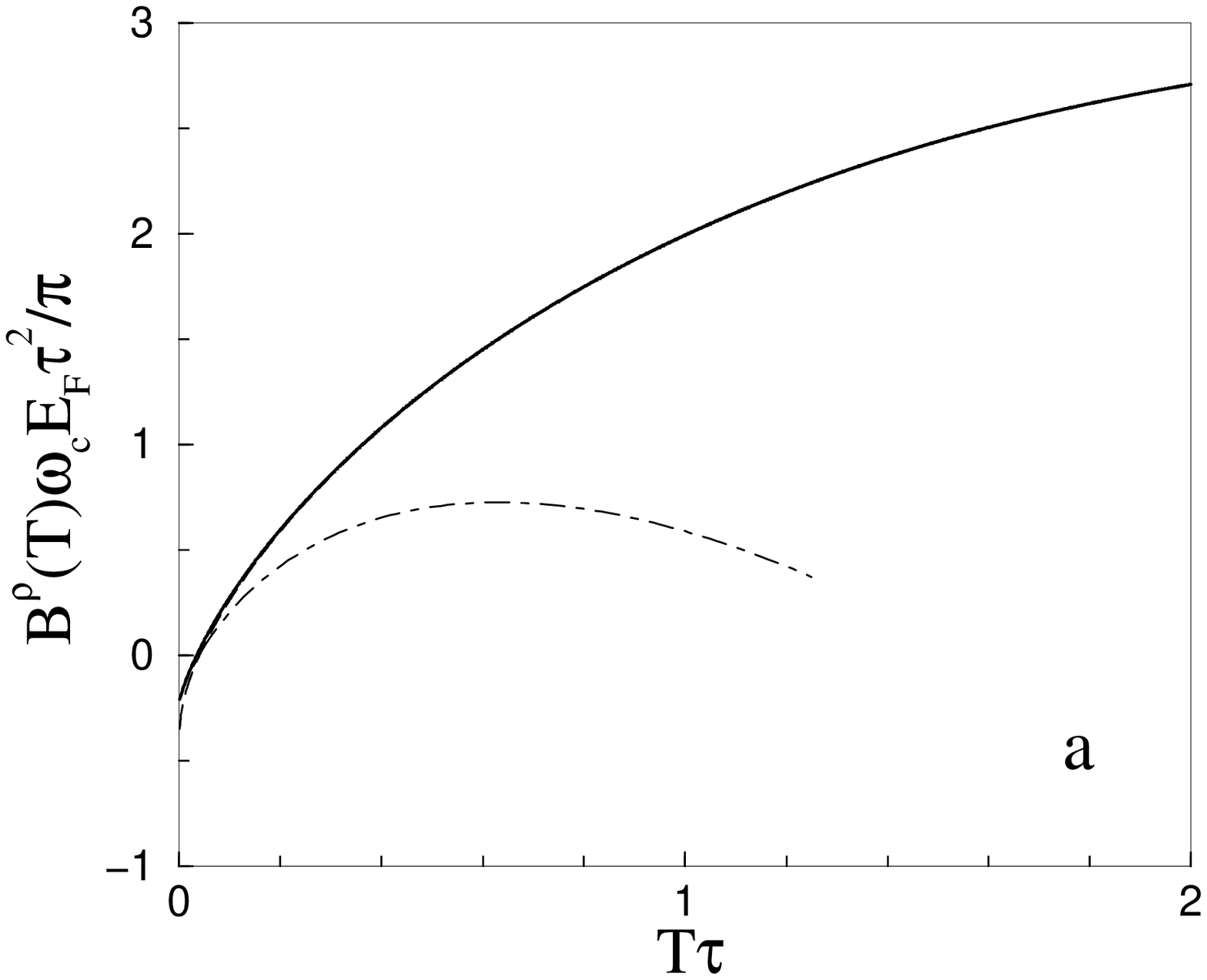} \hskip2cm
 \includegraphics[width=0.35\columnwidth]{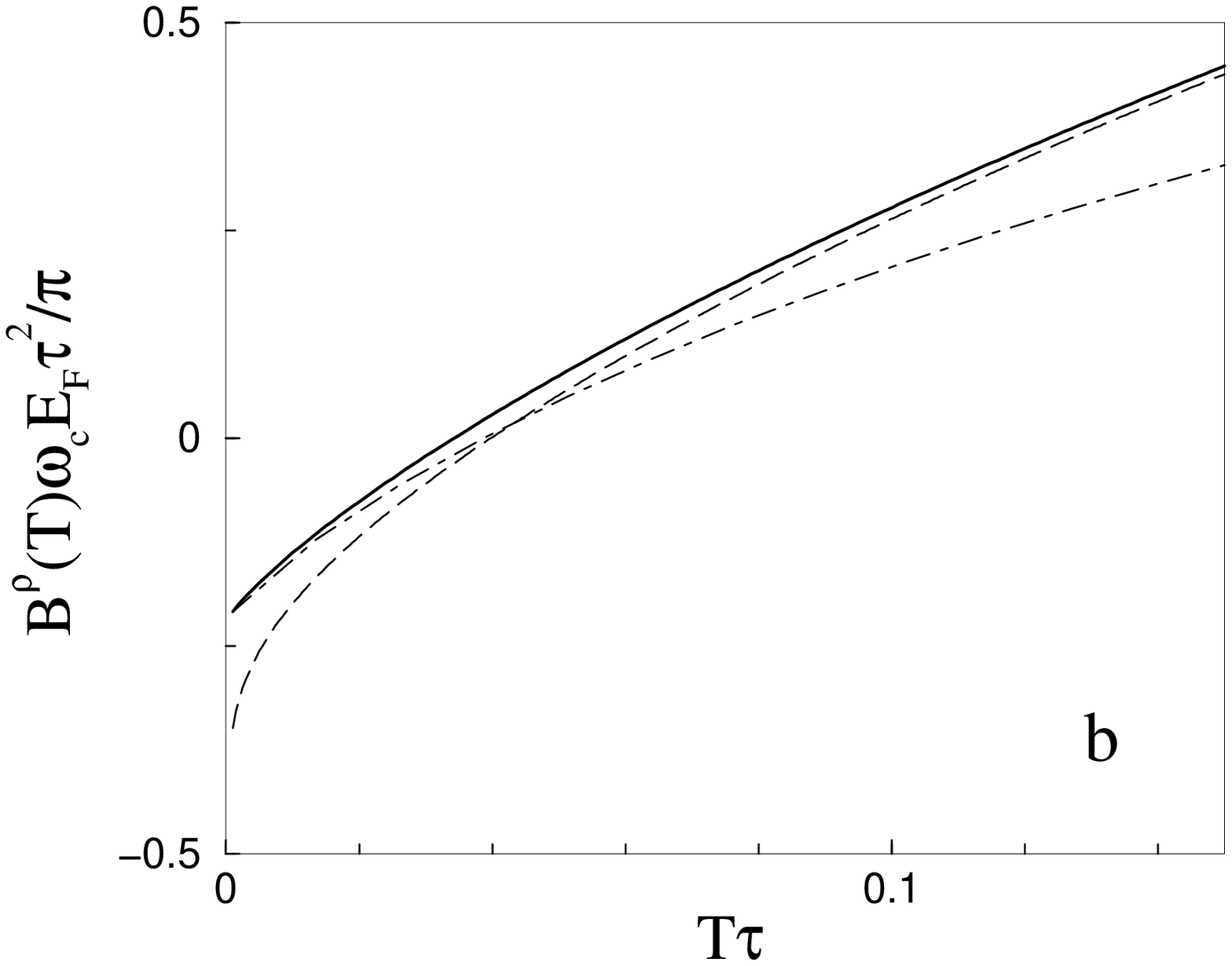}
 \caption{Temperature dependence  of the
 singlet channel correction to the damping factor $B^\rho(T)$
 for $4\pi \ve_F\tau=100$ (solid line)
 with the low-$T$ (dot-dashed) and high-$T$ (dashed) asymptotics.
 (a) Wide temperature range: on this scale $B^\rho(T)$ is essentially indistinguishable from its high-$T$ asymptotics;
 (b) low-$T$ part: the crossover between the two asymptotics occurs at $T\tau\sim 0.05$.}
 \label{Brho}
\end{figure}

\subsection{Coulomb drag in high Landau levels}
Coulomb drag between parallel two-dimensional electron
systems \cite{Eisenstein,Sivan} has developed into a powerful
probe of quantum-Hall
systems \cite{Hill,Rubel,Gramila,Lilly98,kellogg02,kellogg03,Lok,Muraki},
providing information which is complementary to conventional
transport measurements. The drag signal is the voltage $V$ developing
in the open-circuit passive layer
when a current $I$ is applied in the active layer. The drag resistance
(also known as transresistance) is then defined by
$R_D=V/I$.
As a function of interlayer spacing $a$, the interlayer
coupling changes from weak at large
spacings where it can be treated in perturbation theory, to strong at
small spacings where it can result in
states with strong interlayer correlations \cite{kellogg02,kellogg03}.

In a simple picture of Coulomb drag, the carriers of the
active layer transfer momentum to the carriers of the passive layer by
interlayer electron-electron scattering. 
The phase space for interlayer scattering is proportional to the
temperature $T$
in either layer predicting a monotonous temperature dependence
$R_D\propto T^2$ of the drag
resistance. Moreover, the signs of the voltages in active and passive
layer are
expected to be opposite (the same) for carriers of equal (opposite)
charge in the two
layers \cite{MacDonald}. 

Remarkably, experiments show that Coulomb drag behaves very
differently from these simple
expectations  when a perpendicular magnetic field $B$
is applied such that the Fermi energy $\ve_F$
is in a high Landau level, $\ve_F/\hbar\omega_c\gg 1$. ($\omega_c$ is
the cyclotron frequency.)
Several experiments \cite{Gramila,Lok} in the regime of weak interlayer coupling
observed negative drag when the filling factors in the two layers are
different. A more recent experiment \cite{Muraki} also reveals a
non-monotonic dependence on temperature.
While the drag resistivity shows a quadratic temperature dependence at
sufficiently
high temperatures, where drag is always positive, an additional peak
develops at low temperatures
which can have both a positive or a negative sign depending on the
filling-factor difference
between the two layers.

In this Section, we present the theory of Coulomb
drag in the limit of high Landau levels \cite{GMvO}.
In a strong magnetic field, $\omega_c\tau_{\text{tr}}\gg 1$, the intralayer Hall
resistivity $\rho_{xy}$ dominates over the longitudinal resistivity
$\rho_{xx}$. Therefore, the  drag resistivity is given by
\be
\rho_{xx}^D\simeq
\rho_{xy}^{(1)}\ \sigma^D_{yy}\ \rho_{yx}^{(2)}
\label{rhoxx-drag}\ee
\begin{figure}
\centering
\includegraphics[width=0.35\textwidth]{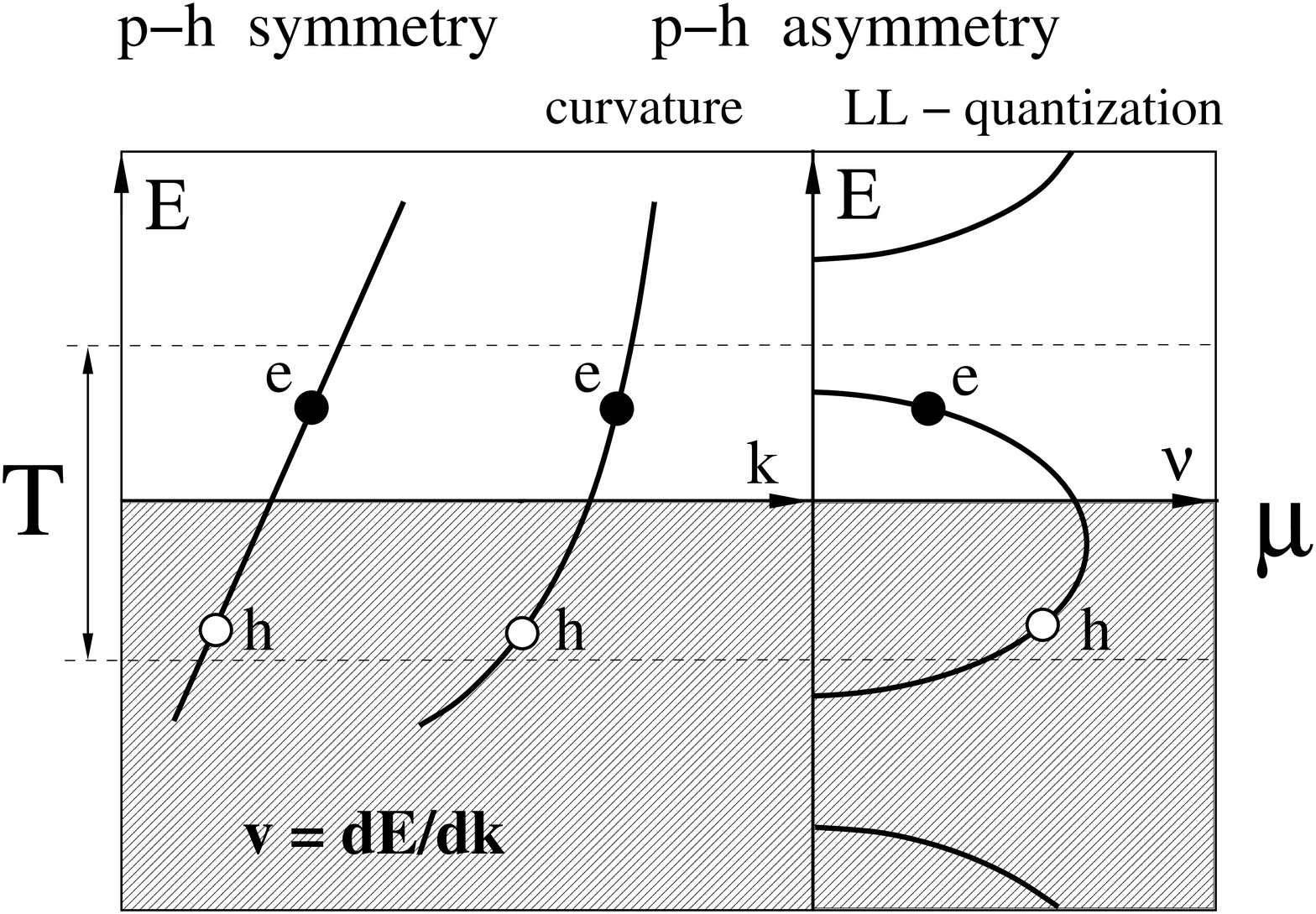}\hskip2cm
\includegraphics[width=0.35\textwidth]{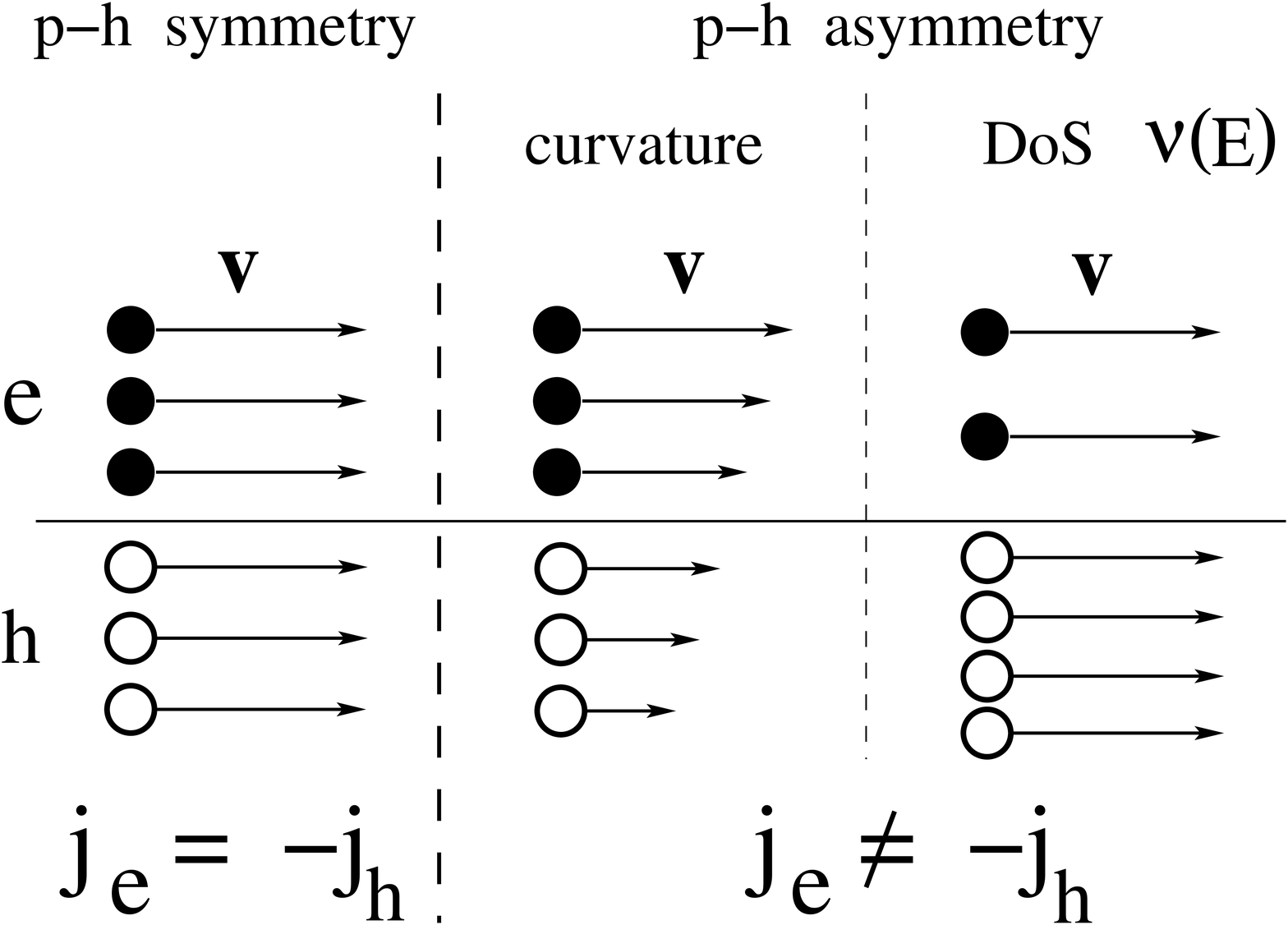}
\caption{Schematic illustration of different sources of particle-hole asymmetry:
curvature of zero-$B$ spectrum $E(k)$ vs LL-quantization of the density of states (DoS)
$\nu(E)$.
In the particle-hole (p-h) symmetric case, the electronic and hole contributions
to the current induced in the passive layer ($j_e$ and
$j_h$, respectively) compensate each other.
When the p-h asymmetry is generated by a finite curvature, the velocities
of electrons and holes (shown by arrows in the right panel) are different,
which destroys the compensation. This is the ``conventional'' mechanism of the drag.
When the DoS depends on energy (in the present case because of the LL-quantization), an
``anomalous'' drag arises due to the difference in numbers of occupied electronic and hole states.
}
\label{fig-ph}
\end{figure}
The Coulomb drag in strong magnetic fields is an interplay
of two contributions, as illustrated in Fig.~\ref{fig-ph}.
At high temperatures, the leading contribution
is due to breaking of particle-hole symmetry by the curvature
of the zero-$B$ electron spectrum. This ``normal'' contribution to the
drag is always positive and increases in a broad temperature range as
$T^2$. At low temperatures, another,  ``anomalous'',
contribution dominates, which arises from the breaking of
particle-hole symmetry by the energy dependence of the density of
states related to Landau quantization. This contribution is sharply
peaked at a temperatute $T\sim\Delta$ (where $\Delta=2\Gamma$ is the Landau
level width) and has an oscillatory sign depending on the density
mismatch between the two layers.

Since the momenta transferred from one layer to the other
are effectively restricted by the inverse interlayer
distance, $a^{-1}$, the behavior of the transresistivity will be
essentially dependent on the relation between $R_c$ and
$a$.
Specifically, with increasing $R_c/a$ the following four regimes are identified
i) diffusive, $R_c/a\ll 1$,
ii) weakly ballistic, $1 \ll R_c/a \ll \omega_c/\Delta$,
iii)   ballistic, $\omega_c/\Delta \ll R_c/a \ll N\Delta/\omega_c$,
and
iv) ultra-ballistic, $N\Delta/\omega_c\ll R_c/a$.
In all regimes, the temperature-dependence of the drag resistivity
is non-monotonous: the absolute value of $\rho_{xx}^D(T)$ shows a peak
around $T\sim \Delta$ and increases again at $T\gg \omega_c.$
However, the $T-$ and $B-$ dependences of $\rho_{xx}^D$, as well
as the sign of the low-temperature peak
(the high-temperature drag is always positive),
are specific for each particular regime, as illustrated in Fig.~\ref{fig7}
and summarized below.

\begin{figure}
\centering
\includegraphics[width=0.95\linewidth]{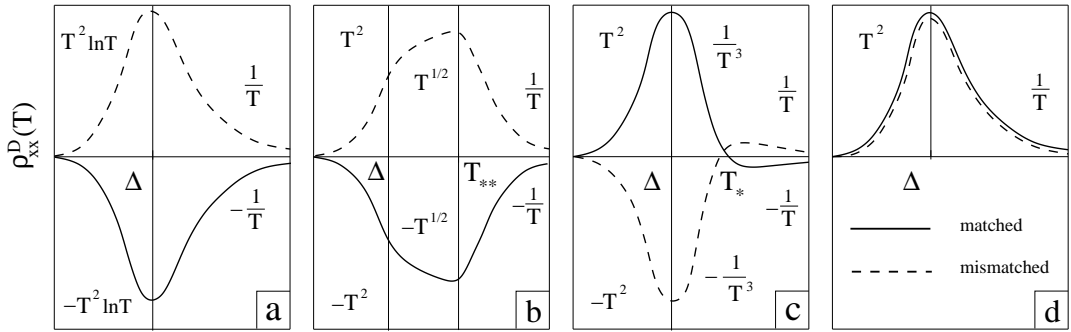}
\caption{Schematic temperature dependence of low-temperature drag in different
regimes:
a) diffusive, $R_c/a\ll 1$;
b) weakly ballistic, $1 \ll R_c/a \ll \omega_c/\Delta$;
c) ballistic, $\omega_c/\Delta \ll R_c/a \ll N\Delta/\omega_c$;
d) ultra-ballistic, $N\Delta/\omega_c\ll R_c/a$.
 }
\label{fig7}
\end{figure}

\textit{Diffusive regime}, $R_c/a\ll 1$.
In the diffusive regime, the drag at not too high temperatures, $T\ll \omega_c,$
is governed by the diffusive rectification\cite{Oppen} which can be calculated 
quasiclassically using the local approximation for the density 
dependence of the conductivity.
As a result, the sign of the drag at $T\sim \Delta$
oscillates but is opposite to what we found above for the
ballistic regime: the drag is negative for equal densities.\cite{Oppen}
At the ``slopes'' of the peak, $\rho_{xx}^D$ scales with $T$ and $B$
in the following way
\begin{equation}
\label{peak-diffusive-scaling}
\rho_{xx}^D \propto \left\{
\begin{array}{ll}
-\ T^2\ \ln (T B^{3/2}), & \qquad \qquad T\ll \Delta, \\[0.2cm]
-\ T^{-1} B^{3/2}\ \ln B, & \qquad \qquad  T\gg \Delta,
\end{array}
\right.
\end{equation}
where the sign corresponds to the case of matching densities.

\textit{Weakly ballistic regime}, $1 \ll R_c/a \ll \omega_c/\Delta$.
This regime is qualitatively similar to the diffusive regime with
\begin{equation}
\label{peak-weakly-ball-scaling}
\rho_{xx}^D \propto \left\{
\begin{array}{ll}
-\ T^2\ B^{-5/4}, & \qquad \qquad T\ll \Delta, \\[0.2cm]
-\ T^{1/2}\ B^{-1/2}, & \qquad \qquad \Delta \ll T \ll T_{**}\equiv\omega_c(a/R_c)\\[0.2cm]
-\ T^{-1}\ B^{5/2}, & \qquad \qquad  T\gg \omega_c(a/R_c),
\end{array}
\right.
\end{equation}
The sign of the peak oscillates just like in the diffusive regime.

\textit{Ballistic regime}, $\omega_c/\Delta\ll R_c/a\ll N\Delta/\omega_c$.
The {\it ballistic} regime is
most relevant experimentally. In this regime, the drag is governed by the 
particle-hole asymmetric effect of Landau quantization of the density of states
and the sign of the drag oscillates,
\begin{equation}
\label{peak-mod-ball-scaling}
\rho_{xx}^D \propto \left\{
\begin{array}{ll}
 T^2\ B\ \ln(B_*/B), & \qquad \qquad T\ll \Delta, \\[0.2cm]
 T^{-3}\ B^{7/2}\ \ln(B_*/B), & \qquad \qquad \Delta \ll T \ll 
 T_*\equiv\Delta\ln^{1/2}({R_c \Delta}/{a \omega_c}),\\[0.2cm]
 -\ T^{-1}\ B^{5/2}, & \qquad \qquad  T\gg T_*,
\end{array}
\right.
\end{equation}
where $B_*\sim (mc/e)(v_F^2/a^2\tau_0)^{1/3}$ (experimentally, the logarithmic
factor in $T_*$ is typically of the order of unity, so that the
intermediate regime may not be fully developed).
We emphasize that the drag at low temperatures is
positive for matched and negative for mismatched densities. 

\textit{Ultra-ballistic regime}, $N\Delta/\omega_c\ll R_c/a$.
The drag for all temperatures is determined by the
conventional contribution related to the curvature of the electron dispersion
 and is always positive,
\begin{equation}
\label{peak-ultra-ball-scaling}
\rho_{xx}^D \propto \left\{
\begin{array}{ll}
T^2\ B^2, & \qquad \qquad T\ll \Delta, \\[0.2cm]
T^{-1}\ B^{7/2}, & \qquad \qquad  T\gg \Delta,
\end{array}
\right.
\end{equation}

At high temperature, $T\gg \omega_c,$ the drag is governed by the conventional
contribution (and is therefore positive) in all the regimes.
It is linear in $T$ in the diffusive regime ($\rho_{xx}^D\propto T B^{-1/2}$).
In all the ballistic regimes the drag resistivity scales as
$\rho_{xx}^D \propto T^2 B^{1/2}$ for $\omega_c \ll T\ll v_F/a$
and $ \rho_{xx}^D \propto T B^{1/2}$ for $ T\gg v_F/a.$

A comparison of Fig.~\ref{fig6} with Fig.~3 of
Ref.~\cite{Muraki} reveals a remarkable agreement between the
experimental findings and the theoretical results. In both the theory
and the experiment,
(i) $\rho_{xx}^D(T)$ shows a sharp peak at low
temperatures;
(ii) the sign of the drag in this temperature range
oscillates as a function of the filling factor of one layer (at fixed
filling factor of the other layer);
(iii) the low-$T$ drag is positive for equal
filling factors and negative when the Fermi energy in one layer is in
the upper half and in the other layer in the lower half of the
Landau band;
(iv) the high-$T$ drag is always positive, independently of the difference
in filling factors of two layers and increases monotonically with increasing $T$.
Furthermore, it was observed by Muraki {\it et al} 
that in the low-temperature regime of initial increase of
$\rho_{xx}^D$, as well as in the high-temperature regime of ``normal''
drag, the drag resistivity can be described by an
empirical scaling law,
$
\rho_{xx}^D\propto(n / B)^{-2.7}f(T/B).
$
Theoretical results for both the low- and high-temperature regimes are in a nice
correspondence with this prediction, with $f(x)\sim x^2$.
\begin{SCfigure}[4]
\centering
\includegraphics[width=0.35\linewidth]{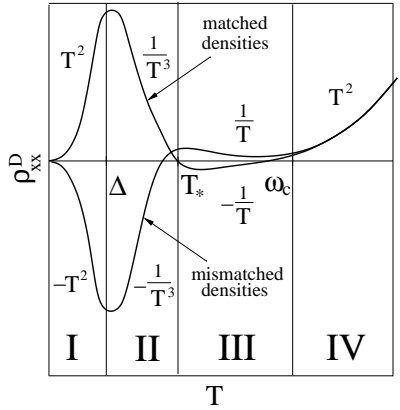} 
\caption{Schematic temperature dependence of drag in the ballistic
  regime for matched and mismatched densities. In the latter case the
  mismatch is chosen such that the drag is negative at low $T$ (see
  text). Scaling of $\rho_{xx}^D$ with temperature in different
  regions is indicated:  $T\ll \Delta$ (I); $\Delta \ll T \ll T_*$ (II);
  $T_*\ll T \ll \omega_c$ (III), and $T\gg \omega_c$ (IV). }
\label{fig6}
\end{SCfigure}

\section{Photoconductivity}
\label{sec5}
\subsection{Microwave-induced magnetoresistance oscillations and zero-resistance states}
Recently, a number of new remarkable effects, important for both
basic and applied physics, have been discovered in two-dimensional
electron systems driven out of equilibrium by strong AC and DC fields.
It was observed \cite{zudov01} that dc resistivity $\rho_{xx}$ of a
high-mobility 2DEG subjected to microwave
radiation of frequency $\omega$ exhibits magnetooscillations with a
period in $\omega$ set by the resonances with multiples of the
cyclotron frequency $\omega_c$. Subsequent work on samples with an
exceptionally high mobility has shown \cite{mani02,zudov03} that for a
sufficiently high radiation power the minima of these
microwave-induced resistance oscillations (MIRO) evolve into ``zero
resistance states''(ZRS), in which the dissipative resistance of a
sample becomes vanishingly small. Unlike oscillatory $\rho_{xx}$, the
Hall resistivity $\rho_{xy}$ remained practically linear in
$\omega_c$. A hallmark of these experimental findings is that the
prominent oscillations of the photoconductivity $\sigma_{\rm ph}\simeq
\rho_{xx}/\rho_{xy}^2$ are observed at magnetic fields as low as
$10\,$mT, and at relatively high temperatures up to $\sim 1\,$K, at
which the Shubnikov-de Haas oscillations are completely suppressed.

Presenting a novel class of magnetooscillations which lead, with
increasing $\omega_c$, to apparently dissipationless transport, the
experimental results \cite{zudov01,mani02,zudov03} have attracted much
theoretical interest. In particular, 
an explanation of the MIRO has been proposed \cite{durst03} in terms of a
combined effect of radiation and Landau quantization on elementary
scattering acts for electrons colliding with impurities (in fact, a
closely related theory was put forward long ago \cite{ryzhii70}). 
A systematic theoretical study of this
mechanism of the MIRO (referred to as a
``displacement'' mechanism in what follows) was carried out in
\cite{vavilov03}. 

On the other hand, it was emphasized \cite{andreev03} that whenever
the linear dc response theory predicts a negative resistivity, this
signifies an instability leading to the formation of domains of
counter-flowing currents. The break-up of an ac-driven sample in
current domains provides an explanation to the experimentally observed
ZRS. 

A different mechanism of the MIRO, called here the ``inelastic''
mechanism, was proposed in \cite{dmitriev03} and studied in more
detail in \cite{short,long} (similar ideas were also discussed in
\cite{dorozhkin03}).  The inelastic mechanism is
associated with a radiation-induced non-equilibrium part of the
distribution function of electrons $f(\ve)$ which oscillates with
varying $\ve\pm\hbar\omega$ due to the Landau quantization. This
mechanism yields the amplitude of oscillations of the linear (with
respect to the dc field) photoconductivity which is proportional to
inelastic scattering time $\tau_{\rm in}$.  The inelastic contribution
dominates over the displacement one for $\tau_{\rm in}$ larger than
single-particle relaxation time $\tau_q$, the condition which is fulfilled in the
experiments.  Apart from the magnitude of the effect, the two
contributions are qualitatively different in their dependence on $T$
and polarization of the radiation. In accord with the experiments, the
inelastic contribution decreases as $\tau_{\rm in}\propto T^{-2}$ with
increasing $T$ and does not depend on the direction of linear
polarization of the microwave field. By contrast, the displacement
mechanism \cite{durst03,ryzhii70,vavilov03} yields a $T$ independent
contribution which depends essentially on the relative orientation of
the microwave and dc fields, which clearly contradicts the
experimental findings.

\subsection{Inelastic mechanism of MIRO}

We consider a high-mobility 2DEG, with $\tau_{\rm q}\ll\tau_{\rm tr}$,
subjected to a classically strong transverse magnetic field,
$\omega_c\tau_{\rm tr}\gg 1$ (we use notations of Sec.~\ref{sec3}, in particular, $\tau_{\rm q}$ and $\tau_{\rm tr}$
are specified in Eq.~(\ref{rates}). The photoconductivity $\sigma_{\rm
  ph}$ determines the longitudinal current flowing in response to a
{\it dc} electric field ${\cal E}_{\rm dc}$, $\vec{j}\cdot\vec{\cal
  E}_{\rm dc}=\sigma_{\rm ph}{\cal E}_{\rm dc}^2$, in the presence of
a microwave electric field ${\bf{\cal E}}_\omega\cos\omega t$. The
more frequently measured
\cite{zudov01,mani02,zudov03,dorozhkin03} longitudinal
resistivity, $\rho_{\rm ph}$, is given by $\rho_{\rm ph}\simeq
\rho_{xy}^2\sigma_{\rm ph}$, where $\rho_{xy}\simeq eB/n_e c$ is the
Hall resistivity, affected only weakly by the radiation.

Here we study the leading inelastic mechanism of MIRO thus taking into
account only effects that are due to a non-trivial energy dependence
of the non-equlibirum distribution function $f(\ve)$.
Photoconductivity $\sigma_{\rm ph}$ is given by the dc responce in the
state with non-equlibirum $f(\ve)$.  According to
Eq.~(\ref{Qdrudedc}), 
\be
\label{photo}
\sigma_{\rm ph}=-\sigma_{\rm dc}^D \textstyle{\int} d\ve \,
\tilde{\nu}^2(\ve)\, \partial_\ve f(\ve),
\ee
where $\tilde{\nu}(\ve)=\nu(\ve)/\nu_0$. The non-equlibirum
distribution function $f(\ve)$ is found as a solution of the
stationary kinetic equation
\be
\label{kineq} {\cal E}^2_\omega\,\frac{\sigma^{\rm
    D}_\omega}{2\omega^2\nu_0}
\sum\limits_{\pm}\tilde{\nu}(\ve\pm\omega)\,[\,f(\ve\pm\omega)-f(\ve)\,]
+\,\,{\cal E}^2_{\rm dc}\,\frac{\sigma^{\rm D}_{\rm
    dc}}{\nu_0\tilde{\nu}(\ve)} \,\frac{\partial}{\partial\ve}
\left[\,\tilde{\nu}^2(\ve)\frac{\partial}{\partial\ve}f(\ve)\,\right]
={f(\ve)-f_T(\ve)\over\tau_{\rm in}}.  \ee On the right-hand side of
Eq.~(\ref{kineq}), inelastic processes are included in the relaxation
time approximation (which is proven \cite{long} to be sufficient under
experimental conditions), and $f_T(\ve)$ is the Fermi distribution.
The left-hand side is due to the electron collisions with impurities
in the presence of the external electric fields.  The first term
describes the absorption and emission of microwave quanta; the rate of
these transitions is proportional \cite{dmitriev03} to $\tau_{\rm
  tr,B}^{-1}(\ve\pm\omega)\propto\nu(\ve\pm\omega)$, see
Eqs.~(\ref{Qdrude}) and (\ref{Qdrude1}). This term can be also
extracted from the kinetic equation of Ref.~\cite{vavilov03}.  The
second term describes the effect of the {\it dc} field and can be
obtained from the first one by taking the limit $\omega\to 0$.
Equation (\ref{kineq}) suggests convenient dimensionless units for the
strength of the {\it ac} and {\it dc} fields:
\be\label{unitsPQ}
{\cal P}_\omega
=\frac{\tau_{\rm in}}{\tau_{\rm tr}}
\left(\frac{e {\cal E}_\omega v_F}{\omega}\right)^2
\frac{\omega_c^2+\omega^2}{(\omega^2-\omega_c^2)^2}~,\qquad
{\cal Q}_{\rm dc}=\frac{2\,\tau_{\rm in}}{\tau_{\rm tr}}
\left(\frac{e {\cal E}_{\rm dc} v_F}{\omega_c}\right)^2
\left(\frac{\pi}{\omega_c}\right)^2~.
\ee
Note that ${\cal P}_\omega$ and ${\cal Q}_{\rm dc}$ are proportional
to $\tau_{\rm in}$ and are infinite in the absence of
inelastic relaxation processes.

To first order in ${\cal P}_\omega$ and ${\cal Q}_{\rm dc}\to 0$,
Eq.~(\ref{kineq}) produces a non-equilibrium correction to $f_T(\ve)$,
$f(\ve)-f_T(\ve)=0.25 {\cal
  P}_\omega\sum_\pm\tilde{\nu}(\ve\pm\omega)\,[\,f_T(\ve\pm\omega)-f_T(\ve)\,]$,
which oscillates both with $\ve/\omega_c$ and $\omega/\omega_c$ due to
$\ve/\omega_c$--oscillations in the DOS. In turn, the oscillatory $f(\ve)$
leads to $\omega/\omega_c$-oscillations of $\sigma_{\rm ph}$,
Eq.~(\ref{photo}),
\be\label{ph1order}
\sigma_{\rm ph}/\sigma_{\rm dc}^D=\langle\,\tilde{\nu}^2(\ve)\,\rangle_\ve
+\,(\omega{\cal P}_\omega/4)\,
\langle\,\tilde{\nu}^2(\ve)
\,\partial_\ve
[\,\tilde{\nu}(\ve+\omega)-\tilde{\nu}(\ve-\omega)\,]
\,\rangle_\ve
\ee
Here we took into account that the SdHO in the experiments are suppressed by
temperature, $T\gg T_D$, so that, analogous to Eq.~(\ref{corr2}), the
energy integration results in averaging over $\ve$ within the period
$\omega_c$, denoted by the angular brackets. For separated LLs,
$\omega_c\tau_{\rm q}\gg 1$, with the semielliptical DOS (\ref{nuSepLLs}),
Eq.~(\ref{ph1order}) gives
\bea
\label{1order}
&&{\sigma_{\rm ph}/\sigma^{\rm D}_{\rm dc}}
=({16\omega_c/3\pi^2\Gamma})  
\left\{1-{\cal P}_\omega ({\omega\omega_c/\Gamma^2})  
\left[\,
\textstyle{\sum}_n \Phi\left({\omega/\Gamma-n\omega_c/\Gamma}\right)
+ O\left({\omega_c {\cal P}_\omega/\Gamma}\right)\,
\right]\, \right\}, \qquad\\
&&4\pi\,\Phi(x)\!=\!x\,{\rm Re}[\, {3\,\rm arccos}(|x|-1)-
(1+|x|)\sqrt{|x|(2-|x|)} \,].
\nonumber
\eea
In the limit of overlapping LLs, the DOS is given by
$\tilde{\nu}=1- 2\delta\cos \frac{2\pi\ve}{\omega_c}$ with
$\delta=\exp(-\pi/\omega_c\tau_{\rm q}) \ll1$. The existence of a
small parameter $\delta$ allows one to calculate $\sigma_{\rm ph}$ to
all orders in ${\cal P}_\omega$ and ${\cal Q}_{\rm dc}$,
\be
\label{result}
\frac{\sigma_{\rm ph}\ }{\sigma^{\rm D}_{\rm dc}}=1+2\delta^2\left[\,1
-
\frac{{\cal P}_\omega\frac{2\pi \omega}{\omega_c}
\sin\frac{2\pi\omega}{\omega_c}
 +4{\cal Q}_{\rm dc}} {1+{\cal P}_\omega
 \sin^2\frac{\pi\omega}{\omega_c}
+{\cal
Q}_{\rm dc}}
\right].
\ee
Results (\ref{result}) and (\ref{1order}) are shown in
Figs.~\ref{fig:phOvLLs} and \ref{fig:phSepLLs} for several values
of ${\cal P}_\omega^{(0)}\equiv{\cal P}_\omega|_{\omega_c=0}$.
\begin{figure}[htb]
\begin{minipage}[t]{.31\textwidth}
\includegraphics[width=\textwidth]{pssb.200743278_Fig19.eps}
\caption{Photoresistivity (normalized to the Drude value) for
overlapping LLs vs $\omega_c/\omega$ at fixed
$\omega\tau_{\rm q}=2\pi$, and for
different ${\cal P}_\omega^{(0)}
=\{0.24,\,0.8,\,2.4\}$. $I-V$ characteristics at the marked
minima are shown in Fig.~\ref{fig:domains}.}
\label{fig:phOvLLs}
\end{minipage}
\hskip3mm
\begin{minipage}[t]{.31\textwidth}
\includegraphics[width=\textwidth]{pssb.200743278_Fig20.eps}
\caption{Photoresistivity (normalized to the Drude value) for
separated Landau levels vs $\omega_c/\omega$ at fixed
$\omega\tau_{\rm q}=16\pi$. The curves correspond to
different levels of microwave power ${\cal P}_\omega^{(0)}
=\{0.01,\,0.03,\,0.05\}$. }
\label{fig:phSepLLs}
\end{minipage}\hskip6mm
\begin{minipage}[t]{.28\textwidth}
\includegraphics[width=\textwidth]{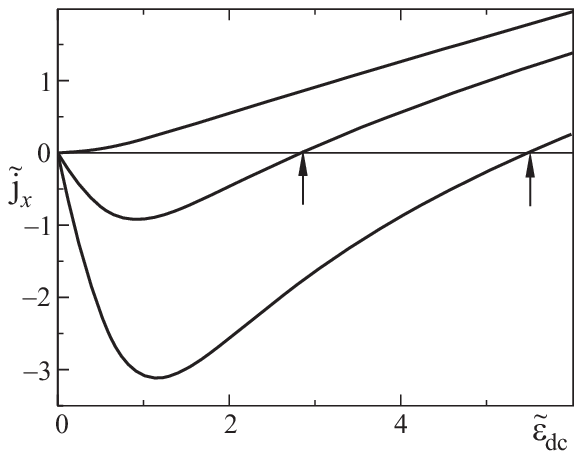}
\caption{Current--voltage characteristics [\,dimensionless current
$\tilde{j}_x=(\sigma_{\rm ph}/\sigma^{\rm D}_{\rm dc}) \tilde{\cal
E}_{\rm dc}$ vs dimensionless field $\tilde{\cal E}_{\rm dc}={\cal
Q}_{\rm dc}^{1/2}$\,] at the points of minima marked by the circles in
Fig.~\ref{fig:phOvLLs}. The arrows show the {\it dc} field $\tilde{\cal E}_{\rm dc}^*$
in spontaneously formed domains.}
\label{fig:domains}
\end{minipage}
\end{figure}
\subsection{Zero-resistance states, strong-field domains, oscillatory compressibility}
At ${\cal P}_\omega$ exceeding certain threshold value ${\cal
  P}_\omega^\star$, $\sigma_{\rm ph}$ around minima becomes negative
(see Figs.~\ref{fig:phOvLLs} and \ref{fig:phSepLLs}). According to
Ref.~\cite{andreev03}, this signifies an instability leading to the
formation of domains with strong Hall field ${\cal E}_{\rm dc}=\pm
{\cal E}_{\rm dc}^\star$ and counter-flowing currents.  The value ${\cal E}_{\rm
  dc}^\star$ is determined by the equation $\sigma_{\rm ph}({\cal
    P}_\omega,{\cal E}_{\rm dc}^\star)=0$, which in the case of overlapping
  LLs, Eq.~(\ref{result}), has the solution (marked by arrows in Fig.~\ref{fig:domains})
\be\label{domains}
{\cal E}_{\rm dc}^*={1\over\sqrt{2}\,\pi}\,
{\omega_c^2\over e v_F}\left({\tau_{\rm
      tr}\over \tau_{\rm in}}\right)^{1/2}
\left[{{\cal P}_\omega\over
{\cal P}_\omega^*}-1\right]^{1/2}~,\qquad{\cal P}_\omega^*=
\left(4\delta^2\frac{\pi\omega}{\omega_c}\sin
\frac{2\pi\omega}{\omega_c}
-\sin^2\frac{\pi\omega}{\omega_c}\right)^{-1}.
\ee
Equation
(\ref{domains}) relates the electric field formed in the domains (measurable
by local voltage probe \cite{willett03}) with the excess power of microwave
radiation. 

In the case of separated LLs, it suffices to keep the linear-in-${\cal
  P}_\omega$ term only even for the microwave power ${\cal
  P}_\omega>{\cal P}^*_\omega\sim\Gamma^2/\omega\omega_c$ at which the
linear-response resistance becomes negative: The second order
correction at ${\cal P}_\omega\sim{\cal P}_\omega^*$ is still small,
$\omega_c {\cal P}_\omega^* /\Gamma \sim \Gamma/\omega \ll 1$.  The
strength of the field in domains is determined by the scale at which
inter-LL elastic scattering becomes efficient, ${\cal E}_{\rm
  dc}^*\sim (\omega_c^2/ e v_F)\sqrt{\tau_{\rm tr}/\tau_{\rm q}}$.

\begin{SCfigure}[4][htb]
\includegraphics[width=.4\textwidth]{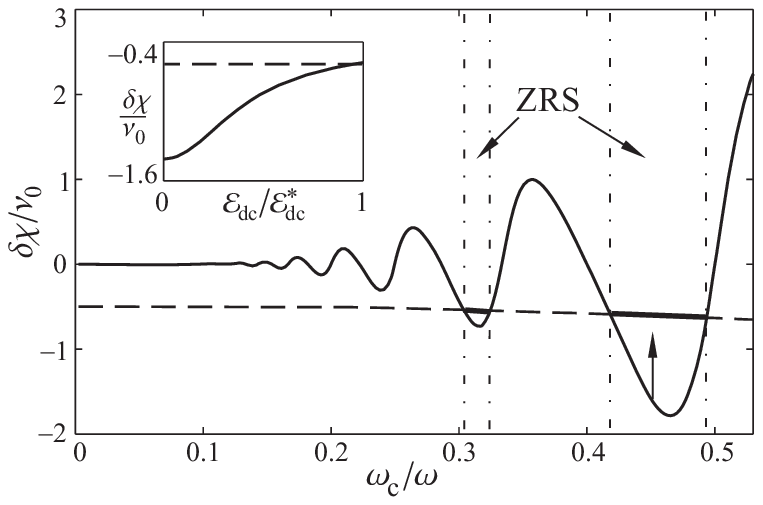}
\caption{The microwave-induced correction to the compressibility
  (solid line) of a 2DEG as a function of $\omega_{\rm c}/\omega$ at
  fixed $\omega\tau_{\rm q}=2\pi$ and microwave power ${\cal
    P}|_{\omega_c=0}=1$. In the zero resistance state (ZRS), the
  electric field inside domains ${\cal E}_{\rm dc}^*$ fixes the
  compressibility at the level shown by a dashed line. Inside the
  domain wall, the electric field ${\cal E}_{\rm dc}$ is smaller than
  ${\cal E}_{\rm dc}^*$ and the compressibility depends on the local
  field as shown in the inset (for $\omega_{\rm c}/\omega=0.45$, this
  ratio is indicated by the arrow. }
\label{fig:compress}
\end{SCfigure}

In Ref.~\cite{compress} it was shown that the local compressibility of an irradiated  2DEG, $\chi=\nu_0+\delta\chi$,
exhibits oscillations similar to the MIRO (see Fig.~\ref{fig:compress}). Calculation using Eqs.~(\ref{kineq}) and (\ref{nuOvLLs}) yields
\be\label{compress}
\delta\chi/\nu_0=\int\!d\ve\,
\tilde{\nu}(\ve) \partial_\ve [f_T(\ve)-f(\ve)]=-\delta^2
\frac{{\cal P}_\omega\frac{2\pi \omega}{\omega_c}
\sin\frac{2\pi\omega}{\omega_c}
 +4{\cal Q}_{\rm dc}} {1+{\cal P}_\omega
 \sin^2\frac{\pi\omega}{\omega_c}
+{\cal
Q}_{\rm dc}}.
\ee
The key features of the effect are: (i) the period and the phase of
the $\omega/\omega_c$-oscillations in $\chi$ are the same as in
$\sigma_{\rm ph}$, Eq.~(\ref{result}); (ii) the amplitude of the
oscillations in $\chi$ and $\sigma_{\rm ph}$ have the same dependence
on the electron temperature and microwave power; (iii) the ZRS
corresponds to a plateau in the compressibility: inside the domains
$\chi=\nu_0(1-2\delta^2)/2$.  Local measurements of the
compressibility may provide a real space snapshot of the domain
structure in the ZRS.  Experimental work in this direction is
currently underway.

\subsection{High-power effects, subleading mechanisms, fractional MIRO}

In addition to the peak-valley structure near integer $\omega/\omega_c$ (integer
MIRO), several experiments \cite{zudov01,zudov03,dorozhkin03,zudov04,%
  dorozhkinINTRA, multi,dorozhkin06} reported similar features near
certain fractional values, $\omega/\omega_c=1/2,\,3/2,\,5/2,\,2/3..$
(``fractional MIRO'', or FMIRO), which at elevated microwave power
also evolved into ZRS (``fractional ZRS'') \cite{multi}. Initially,
FMIRO were ascribed to multiphoton
processes \cite{zudov04,leiliumulti}. This expanation, however, failed
to reproduce the observations \cite{dorozhkin06}, where
the FMIRO only occured at $\omega$ below a certain threshold value. It was
shown that the threshold can be explained in the framework of the
single-photon inelastic mechanism \cite{dorozhkin06,crossover}. Here, the
FMIRO near combined resonances $n\omega=m\omega_c$ occured due to a resonant series of
$n$ single-photon transitions with {\it real} absorption (emission) of
the microwave quanta \cite{crossover}, in distinction to the virtual
multiphoton processes. A systematic theory of the FMIRO \cite{separated} has shown that the existing
theories \cite{leiliumulti, crossover} miss several important
contributions. In particular, in the limit of well separated LLs the
FMIRO are dominated by multiphoton inelastic mechanism. 
Provided $\tau_{\rm in}/\tau_{\rm q}\gg1$, the multiphoton displacement mechanism
\cite{leiliumulti} yields a parametrically smaller contribution and can be neglected.
At weaker magnetic field the effects related to microwave-induced
sidebands in the DOS become important. Close to the magnetic
field at which the LLs start to overlap, the FMIRO are dominated by
the single-photon inelastic mechanism \cite{crossover}. Finally, in the
regime of strongly overlapping LLs the FMIRO get exponentially suppressed.

A unified picture of the photoresponce in the limit of overlapping LLs
was recently developed in \cite{verylong}. On top of nonlinear
interplay between the inelastic and displacement mechanisms at
elevated microwave power, two novel mechanisms leading to the MIRO,
``quadrupole'' and ``photovoltaic'', were identified. In the
quadrupole mechanism, the microwave radiation leads to excitation of
the second angular harmonic of the distribution function. The dc
response in the resulting nonequilibrium state yields an oscillatory
contribution to the Hall part of the photoconductivity tensor which
violates Onsager symmetry. In the photovoltaic mechanism, a combined
action of the microwave and dc fields produces non-zero temporal
harmonics of the stationary distribution function. The ac response in
this state contributes to both the longitudinal and Hall MIRO.
Provided $\tau_{\rm in}/\tau_{\rm q}\gg 1$, the inelastic mechanism
still gives the dominant contribution to the diagonal part of the
photoconductivity tensor. However, the quadrupole and photovoltaic
mechanisms are the only ones yielding oscillatory corrections to the
Hall part. Further, it was shown that a competition between various
nonlinear effects (the feedback effects, the excitation of high
angular and temporal harmonics of the distribution function, and the
multiphoton effects) drives the system through four different
nonlinear regimes with increasing microwave power. Most dramatic
changes in the photoresponse are due to the feedback effects. At
${\cal P}_\omega\gg1$, the feedback from the microwave--induced
oscillations of the isotropic part of the distribution $f(\ve)$ leads
to the saturation of the inelastic contribution, and
to the strong interplay of the inelastic effect and all other
contributions to the MIRO. In particular, the strong oscillations of
$f(\ve)$ change sign of the most relevant parts of the displacement
and photovoltaic contributions.  At higher power, ${\cal
  P}_\omega\gg\omega_c\tau_{\rm in}$, the feedback suppresses the
effects on higher temporal and angular harmonics of the distribution
function. At still higher power, ${\cal P}_\omega\gg\tau_{\rm
  in}/\tau_{\rm q}$, the multiphoton excitation becomes pronounced and
starts to compete with the feedback effects. Finally, at ${\cal
  P}_\omega\gg\tau_{\rm in}/\omega_c^2\tau_{\rm q}^3$, the feedback and
multiphoton effects destroy all quantum contributions, restoring the
classical Drude conductivity.

\subsection{New developments and open questions}

In spite of the essential advances in the understanding of nonequilibrium
magnetotransport phenomena in a 2DEG, a number of questions remain open.
A puzzling insensitivity of the MIRO to the direction of circular polarization of
the microwave field was reported in \cite{polarization}. The strong
interplay \cite{ACDC} between the dc-- \cite{yang02,2componentDC} and
microwave--induced oscillations deserves theoretical study. A further 
challenging 
direction of future experimental and theoretical research is the transition to
the ZRS, the domain structure, electron transport and noise in the ZRS; first
steps in this direction have been made in \cite{Halperin,Balents}. In
particular, the theory has not explained the seemingly activated temperature
dependence of the residual resistance in the ZRS.  Recently discovered
$B$-periodic magnetooscillations \cite{B-periodic} 
which were ascribed to the microwave excitation of 2D edge
magnetoplasmons warrant further investigation; in particular, their
microscopic mechanism is still unclear.

\begin{acknowledgement}
 Most of the work reported in this review has been done in the
 framework of the Schwerpunktprogramm ``Quanten-Hall-Systeme'' of the
 Deutsche Forschungsgemeinschaft. We acknowledge collaboration with
 Y. Adamov, I.L. Aleiner, S.I. Dorozhkin, O. Entin-Wohlman, Y. Levinson,
 I.V. Pechenezhskii, E. Tsitsishvili, M.G. Vavilov, F. von Oppen, and J. Wilke
 on topics reviewed in this article.
 We are grateful to P.T. Coleridge, W. Dietsche, S.I. Dorozhkin, R.R. Du, 
 A.V. Germanenko, M.E. Gershenson, R. Haug, 
 I.V. Kukushkin, Z.D. Kvon, J.G.S. Lok, R.G. Mani, G.M. Minkov,
 C. Mitzkus, E.B. Olshanetsky, V.M. Pudalov, V. Renard, 
 A.K. Savchenko, J.H. Smet, K. von Klitzing, D. Weiss, and
 M.A. Zudov for informing us on the experimental results and for 
 stimulating discussions.
 We also acknowledge support by the
 DFG Center for Functional Nanostructures, by EUROHORCS/ESF (I.V.G.), 
 as well as by the INTAS Project 05-1000008-8044.  
 
\end{acknowledgement}


\begin{thebibliography}{10}
\bibitem{footAndo} The consideration of ultra-long-range disorder with $d\gg l_B$ in
  \cite{ando} is not justified since this condition violates the
applicability of SCBA.
\end{thebibliography}
\end{document}